\documentclass[structabstract]{aa}
\usepackage{txfonts}
\usepackage{natbib}
\usepackage{graphicx}
\usepackage{aalongtable}
\bibpunct{(}{)}{;}{a}{}{,}

\begin{document}

\title{KIC\,10526294: a slowly rotating B star with rotationally split, quasi-equally spaced gravity modes\thanks{Based on observations made with the William Herschel Telescope operated by the Isaac Newton Group on the island of La Palma at the Spanish Observatorio del Roque de los Muchachos of the Instituto de Astrof\'{i}sica de Canarias.}}

\author{P.~I.~P\'{a}pics\inst{\ref{inst1}}
\and E.~Moravveji\inst{\ref{inst1}}\thanks{Postdoctoral Fellow of the Belgian Science Policy Office (BELSPO), Belgium}
\and C.~Aerts\inst{\ref{inst1},\ref{inst2}}
\and A.~Tkachenko\inst{\ref{inst1}}\thanks{Postdoctoral Fellow of the Fund for Scientific Research (FWO), Flanders, Belgium}
\and S.~A.~Triana\inst{\ref{inst1}}\thanks{Postdoctoral Fellow of the Fund for Scientific Research (FWO), Flanders, Belgium}
\and S.~Bloemen\inst{\ref{inst1},\ref{inst2}}
\and J.~Southworth\inst{\ref{inst3}}}

\institute{Instituut voor Sterrenkunde, KU Leuven, Celestijnenlaan 200D, B-3001 Leuven, Belgium\\ \email{Peter.Papics@ster.kuleuven.be}\label{inst1}
\and Department of Astrophysics, IMAPP, Radboud University Nijmegen, PO Box 9010, 6500 GL Nijmegen, The Netherlands\label{inst2}
\and Astrophysics Group, Keele University, Staffordshire, ST5 5BG, United Kingdom\label{inst3}}

\date{Received 29 April 2014 / Accepted 1 July 2014}

\abstract{Massive stars are important for the chemical enrichment of the universe. Since internal mixing processes influence their lives, it is very important to place constraints on the corresponding physical parameters, such as core overshooting and the internal rotation profile, so as to calibrate their stellar structure and evolution models. Although asteroseismology has been shown to be able to deliver the most precise constraints so far, the number of detailed seismic studies delivering quantitative results is limited.}{Our goal is to extend this limited sample with an in-depth case study and provide a well-constrained set of asteroseismic parameters, contributing to the ongoing mapping efforts of the instability strips of the $\beta$\,Cep and slowly pulsating B (SPB) stars.}{We derived fundamental parameters from high-resolution spectra using spectral synthesis techniques. We used custom masks to obtain optimal light curves from the original pixel level data from the \textit{Kepler} satellite. We used standard time-series analysis tools to construct a set of significant pulsation modes that provide the basis for the seismic analysis carried out afterwards.}{We find that KIC\,10526294 is a cool SPB star, one of the slowest rotators ever found. Despite this, the length of \textit{Kepler} observations is sufficient to resolve narrow rotationally split multiplets for each of its nineteen quasi-equally spaced dipole modes. The number of detected consecutive (in radial order) dipole modes in this series is higher than ever before. The observed amount of splitting shows an increasing trend towards longer periods, which -- largely independent of the seismically calibrated stellar models -- points towards a non-rigid internal rotation profile. From the average splitting we deduce a rotation period of $\sim188\,\mathrm{days}$. From seismic modelling, we find that the star is young with a central hydrogen mass fraction $X_c>0.64$; it has a core overshooting $\alpha_{ov}\le0.15$.}{}

\keywords{Asteroseismology - 
Stars: variables: general - 
Stars: early-type -
Stars: fundamental parameters - 
Stars: oscillations - 
Stars: rotation}

\titlerunning{KIC\,10526294: rotational splitting of gravity modes in a slowly rotating B star}
\maketitle


\section{Introduction}\label{intro}

The \textit{Kepler} satellite is the latest iteration in the line of the recent space missions after MOST \citep[Microvariablity and Oscillations of STars,][]{2003PASP..115.1023W} and CoRoT \citep[Convection Rotation and planetary Transits,][]{2009A&A...506..411A}, providing a virtually uninterrupted photometric monitoring of more than 150\,000 stars with micromagnitude-precision in a 105-square-degree field of view (FOV) fixed between the constellations of Cygnus and Lyra. Although the primary goal of the mission is to detect Earth-like exoplanets \citep{2010Sci...327..977B}, the assembled data set has a huge potential for asteroseismology, the study of stellar interiors through the detection and interpretation of pulsation modes \citep{2010PASP..122..131G}.

Massive stars play a dominant role in the chemical evolution of the Universe. Since their structure on the main sequence is dominated by a convective core and a radiative envelope, these stars harbour internal mixing processes, which have a significant influence on their lifetime by enhancing the size of the convective region where mixing of chemical elements occurs. Efficient mixing might be induced by the presence of convective core overshooting, for instance, but also by a non-rigid internal rotation profile \citep[see, e.g.,][]{2013sse..book.....K}. Despite their importance, the physical parameters describing these processes are hardly known, and better observational constraints are necessary to achieve a precision that enables us to calibrate the stellar structure and evolution models of massive stars.

B-type stars on the main sequence are not amongst the most massive stars, but they all share this structure. Since many of them show non-radial oscillations \citep[such as the $\beta$\,Cep pulsators, the slowly pulsating B -- or SPB -- stars, and some Be stars, see, e.g.,][]{2010aste.book.....A}, they are the perfect targets for seismic studies, which have already been shown to carry the potential of delivering constraints on the amount of core overshooting \citep[][]{2003Sci...300.1926A} and on the internal rotation profile \citep[see, e.g.,][]{2004A&A...415..241A,2004A&A...415..251D}. Despite a few further modelling efforts (based on extensive ground based multi-site campaigns involving multi-colour photometry and high-resolution, high S/N spectroscopy), the number of stars for which core overshooting and the internal rotation profile was constrained in practice remains very limited, and the constraints have limited precision \citep{2013EAS....64..323A}.

Observations from space -- especially the analysis of CoRoT data -- has revealed a much wider diversity in the photometric variability of main sequence B-type stars than expected \citep[for an overview, see, e.g.,][]{papics_phd}. The diversity in the behaviour of  B-type stars on the main sequence must imply that the details in the internal physics of these various stars must be different. In the context of this paper, it is important to recall the detection of series of $g$ modes (consecutive in radial order, having the same $\ell$ value) with nearly equidistant spacings in period space for the hybrid SPB/$\beta$\,Cep pulsators HD\,50230 \citep{2010Natur.464..259D,2012A&A..542A..88D} and HD\,43317 \citep{2012A&A...542A..55P}. These series, and the deviations from the uniform spacing, carry information about the extent of the fully mixed convective core, and the extra mixing processes operating in the radiative region just above the core \citep[see, e.g.,][]{2008MNRAS.386.1487M}, thus they are viable tools for deducing the size of the core, the overshooting parameter, and the near-core mixing processes.

KIC\,10526294 is one of the eight main sequence B-type stars included in our \textit{Kepler} guest observer (GO) proposals (Cycle 3-4). We have already presented the in-depth analysis of the two double-lined binaries of the sample \citep{2013A&A...553A.127P} -- the first detailed analysis of pulsating main sequence B-type stars based on several years of \textit{Kepler} photometry and high resolution spectroscopy -- while in this paper, we discuss the single B star, which seems to be the most promising case for asteroseismology.


\section{Fundamental parameters}\label{spectroscopy}
KIC\,10526294 is a single star in the constellation of Lyra. Given its low apparent brightness (\textit{Kepler} magnitude of 13.033), no studies have been made about it so far, and there is also no parallax measurement available. The known astrometric and photometric parameters are listed in Table\,\ref{apriori}. It was identified as an SPB candidate independently by \citet{2011A&A...529A..89D} and \citet{2012AJ....143..101M} directly from \textit{Kepler} light curves (using an automated supervised classification of public Q1 data, and 2MASS colours to distinguish between the SPB and $\gamma$\,Dor groups that have similar asteroseismic properties) or with data mining techniques involving a combination of existing catalogues (multicolour photometry, proper motions, etc.), respectively. This classification was not confirmed by spectroscopy until our investigation (see Sect.\,\ref{spectroscopy}). The star was not included in the B-type sample assembled by \citet{2011MNRAS.413.2403B}.

\begin{table}
\caption{Basic observational properties of KIC\,10526294.}
\label{apriori}
\centering
\renewcommand{\arraystretch}{1.25}
\setlength{\tabcolsep}{0pt}
\begin{tabular}{l@{\hskip 12pt} r l@{\hskip 1pt} c@{\hskip 1pt} l r}
\hline\hline
Parameter && 			\multicolumn{3}{l}{KIC\,10526294} & Ref.\\
\hline
$\alpha_{2000}$&&		\multicolumn{3}{l@{\hskip 12pt}}{$\mathrm{ }19^\mathrm{h}12^\mathrm{m}02\fs738$}&1\\
$\delta_{2000}$&		$+$&\multicolumn{3}{l@{\hskip 12pt}}{$47\degr42\arcmin36\farcs46$}&1\\
\textit{Kepler} mag.&&	$13.033$&&&1\\
$\textit{g}_\mathrm{SDSS}$&&$12.834$&&&1\\
$\textit{r}_\mathrm{SDSS}$&&$13.099$&&&1\\
$\textit{i}_\mathrm{SDSS}$&&$13.342$&&&1\\
$\textit{z}_\mathrm{SDSS}$&&$13.483$&&&1\\
2MASS ID&&				\multicolumn{3}{l@{\hskip 12pt}}{J19120273+4742364}&2\\
$J_\mathrm{2MASS}$&&		$12.945$&$\pm$&$0.025$&2\\
$H_\mathrm{2MASS}$&&		$13.020$&$\pm$&$0.021$&2\\
$K_\mathrm{2MASS}$&&		$13.045$&$\pm$&$0.026$&2\\
USNO-B1.0 ID&&				\multicolumn{3}{l@{\hskip 12pt}}{1377-0396062	}&3\\
$B\textit{1}_\mathrm{USNO}$&&		$13.36$&&&3\\
$R\textit{1}_\mathrm{USNO}$&&		$13.73$&&&3\\
\hline
\end{tabular}
\tablebib{(1) \citet{2009yCat.5133....0K}; (2) \citet{2mass}; (3) \citet{2003AJ....125..984M}.}
\end{table}

To confirm that the stars identified as SPB pulsators and included in our \textit{Kepler} GO sample are indeed main sequence B-type stars, we have taken spectra of all targets. KIC\,10526294 was observed using the the ISIS spectrograph mounted on the 4.2-metre William Herschel Telescope on La Palma (Spain). The resolving power of the blue ($4169-4571\,\AA$) and red ($6037-6830\,\AA$) arms of the instrument in the chosen observing mode was $R \approx 22\,000$ and $13\,750$, respectively. Two consecutive exposures were taken on 11 June 2012 using integration times of 400\,s and 1200\,s. The raw frames were reduced using standard STARLINK routines following the optimal extraction described by \citet{1989PASP..101.1032M}. We only considered the second exposure -- having a significantly higher signal-to-noise ratio ($\mathrm{S/N} \approx 42$ and $61$ for the blue and red arm, measured in the line free regions of $4200-4230\,\AA$ and $6160-6225\,\AA$, respectively) -- for our analysis.

The reduced spectrum was rectified using an interactive graphical user interface (GUI). This was done with cubic splines that were fitted through a few tens of points at fixed wavelengths, where the continuum was known to be free of spectral lines. The GUI helps in this process by displaying a synthetic spectrum (built with a predefined set of fundamental parameters, which are possible to adjust on-the-fly) in the background, providing a quick feedback about the position of line-free regions and about the goodness of the rectification.

We used the GSSP package \citep{2011A&A...526A.124L,2012MNRAS.422.2960T} to derive the fundamental parameters from the normalised spectrum. This program compares observed and synthetic spectra computed in a grid of $T_\mathrm{eff}$, $\log g$, $\xi_\mathrm{t}$, $[M/H]$, and $v \sin i$, and provides the optimum values of these parameters for the fit where a minimum in $\chi^2$ is reached. In the next step, individual abundances of different chemical species can be adjusted assuming a stellar atmosphere model with a given overall metallicity, but given the relatively low S/N and wavelength coverage, we did not derive these values. The error bars are represented by $1$-$\sigma$ confidence levels that were computed from $\chi^2$ statistics. The chosen grid of atmosphere models was computed using the most recent version of the LLmodels code \citep{2004A&A...428..993S}. For the calculation of synthetic spectra, we used the LTE-based code SynthV \citep{1996ASPC..108..198T}, which allows the spectra to be computed based on individual elemental abundances if necessary.

\begin{figure*}
\resizebox{\hsize}{!}{\includegraphics{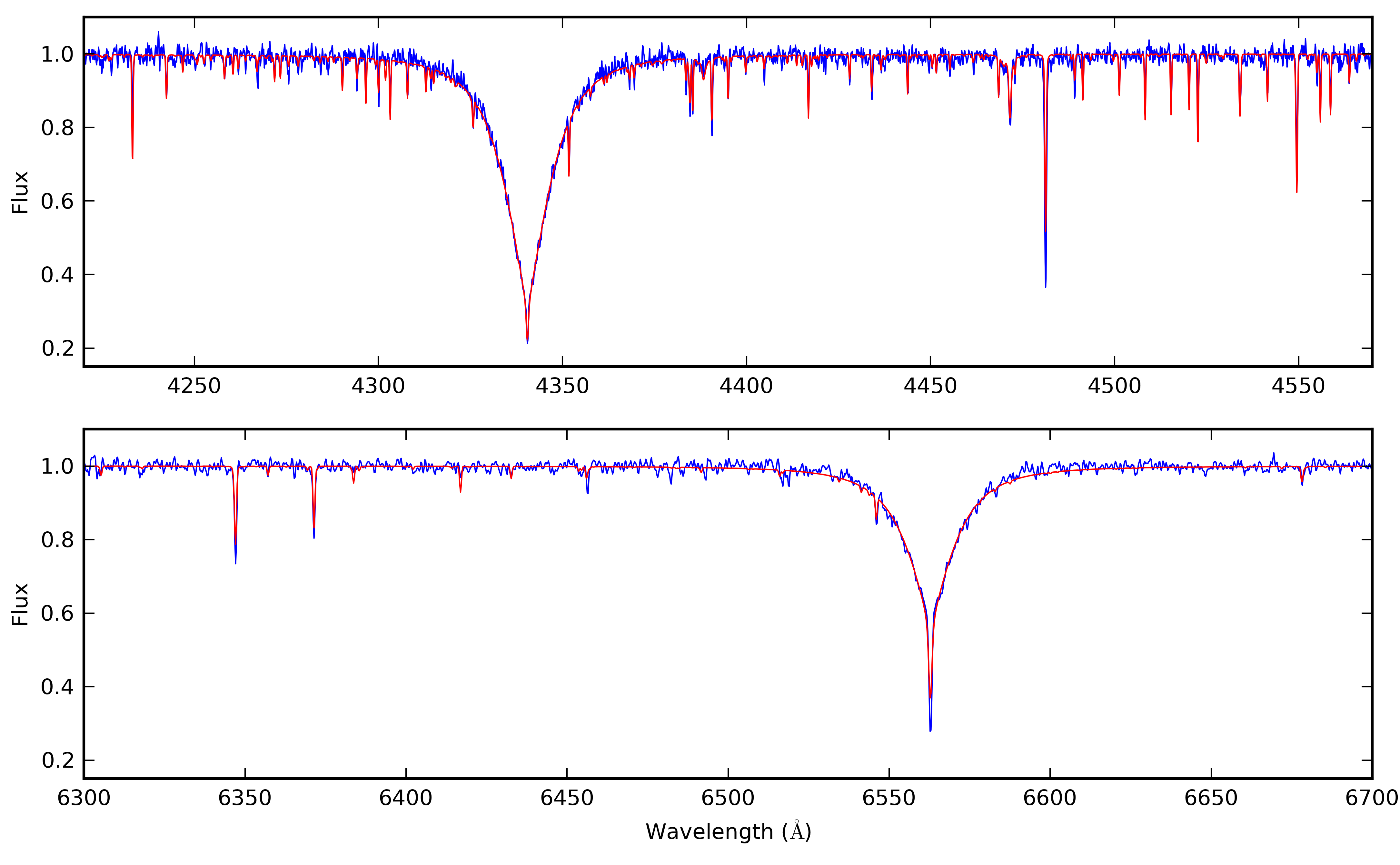}} 
\caption{Comparison of rectified observed, and synthetic spectra for two selected wavelength regions of KIC\,10526294. In each panel, the observed ISIS spectrum is plotted with a blue solid line, and the synthetic spectrum is plotted with a red solid line.}
\label{spectralfit}
\end{figure*}

\begin{figure}
\resizebox{\hsize}{!}{\includegraphics{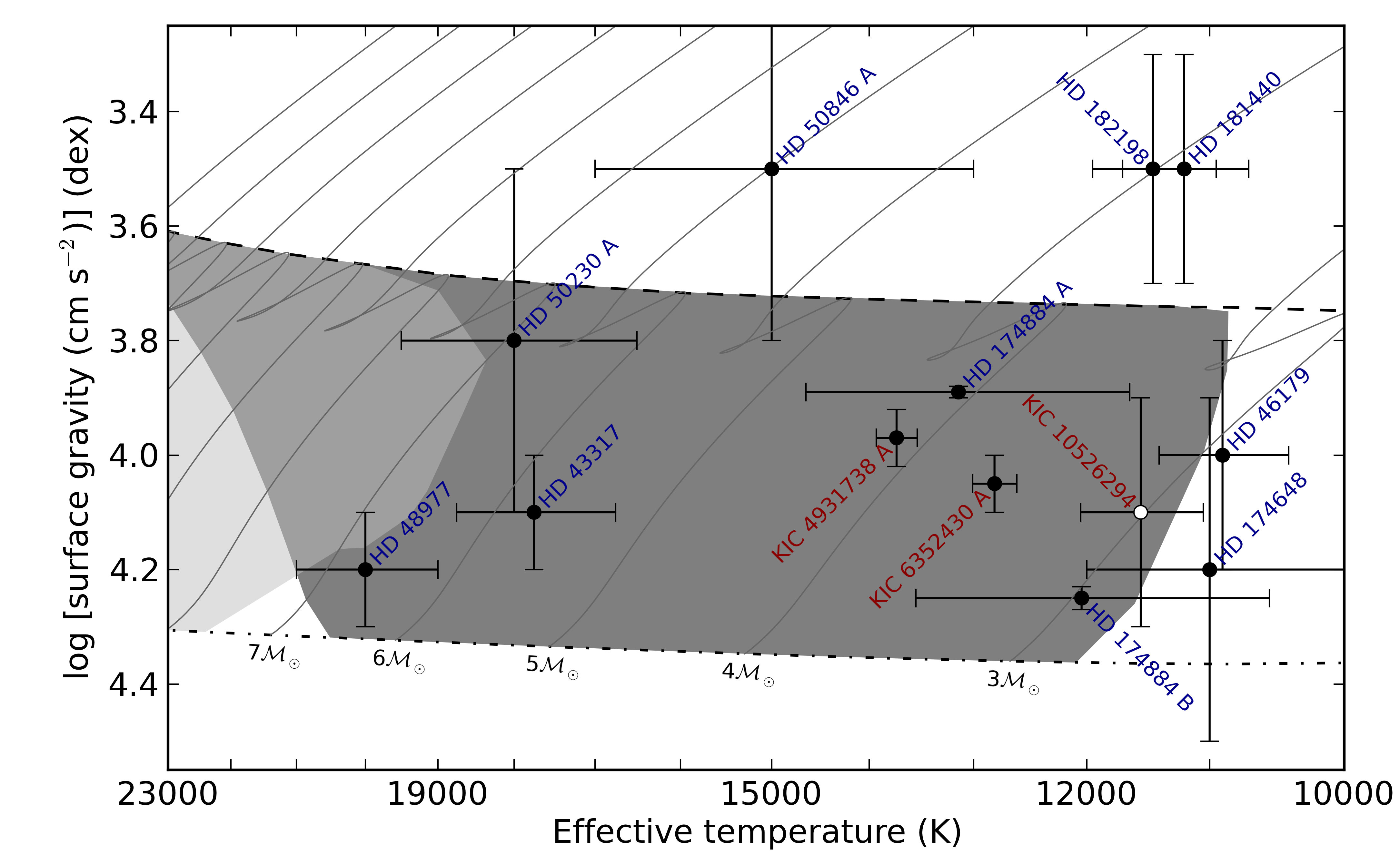}}
\caption{Kiel diagram ($\log T_\mathrm{eff}$ versus $\log g$) of a sample of B-type stars near the main sequence -- for which an in-depth analysis was carried out -- observed by CoRoT (labelled in dark blue) and by \textit{Kepler} (labelled in dark red), showing the position of KIC\,10526294 (plotted with a white marker). The dot-dashed line represents the zero-age main sequence (ZAMS), while the dashed line represents the terminal-age main sequence (TAMS). The thin grey lines denote evolutionary tracks for selected masses, and the $\beta$\,Cep (light grey) and SPB (dark grey) instability strips \citep{1977AcA....27...95D} are also plotted. These were all calculated for $Z = 0.015$, $X=0.7$, using OP opacities \citep{2005MNRAS.362L...1S} and A04 heavy element mixture \citep{2005ASPC..336...25A}. Other details are the same as in \citet{1999AcA....49..119P}. Different error bars reflect differences in data and methodology.}
\label{hrdiagram}
\end{figure}

Table\,\ref{fundparams} lists the atmospheric parameters for KIC\,10526294. The spectral type and the luminosity class have been derived by interpolating in the tables published by \citet{1982SchmidtKalerBook}. Figure\,\ref{spectralfit} shows observed spectrum and the synthetic fit, and Fig.\,\ref{hrdiagram} shows the position of KIC\,10526294 in the Kiel-diagram, situated well within the theoretical SPB instability strip \citep{1999AcA....49..119P}.

\begin{table}
\caption{Fundamental parameters of KIC\,10526294.}
\label{fundparams}
\centering
\renewcommand{\arraystretch}{1.25}
\setlength{\tabcolsep}{1pt}
\begin{tabular}{l l@{\hskip 60pt} r c l c r c l}
\hline\hline
\multicolumn{2}{l}{Parameter} & \multicolumn{3}{c}{KIC\,10526294}\\
\hline
\multicolumn{2}{l}{$T_\mathrm{eff}\,(\mathrm{K})$} & $11550$&$\pm$&$500$\\
\multicolumn{2}{l}{$\log g\,\mathrm{(cgs)}$} & $4.1$&$\pm$&$0.2$\\
\multicolumn{2}{l}{$Z$} & $0.016$&$\pm$&$^{0.013}_{0.007}$\\
\multicolumn{2}{l}{Gaussian line broadening\,$(\mathrm{km\,s}^{-1})$} & $18$&$\pm$&$4$\\
\multicolumn{2}{l}{$\xi_\mathrm{t}\,(\mathrm{km\,s}^{-1})$}& $2.0$&\multicolumn{2}{l}{(fixed)}\\
\multicolumn{2}{l}{Spectral type\tablefootmark{a}}&\multicolumn{3}{c}{B8.3\,V}\\
\hline
\end{tabular}
\tablefoot{\tablefoottext{a}{Spectral type have been determined based on $T_\mathrm{eff}$ and $\log g$ values by using an interpolation in the tables given by \citet{1982SchmidtKalerBook}.}}
\end{table}

\section{The \textit{Kepler} light curve}
\subsection{Description of the data}
Owing to the mission design (Earth-trailing heliocentric orbit, fixed field of view, data downlink, and rolling the spacecraft by 90 degrees every quarter year to reposition the solar arrays and the heat radiators) data from the \textit{Kepler} satellite is delivered in quarters, with Q0 being the commissioning run, Q1 the first (although not full-length) science quarter, etc. Each quarterly roll and downlink period results in a small gap in the data, and with each repositioning, targets move from one module to the next on the CCD array. This means that to construct a full, continuous light curve one needs to correct for the slight differences between the characteristics (e.g., sensitivity) of the modules in question. Furthermore, long-term trends might arise from slight shifts in the point spread functions' (PSF) position over the frame when the standard pixel mask is too small. Although this is a step from optimal S/N to optimal signal (since we include pixels with less signal, thus a lower S/N level), extending the coverage of these masks (making sure no nearby stars are included) provides better long-term stability for the quarterly light curves \citep[for further explanation and examples, see][]{2013A&A...553A.127P}.

\begin{figure}
\resizebox{\hsize}{!}{\includegraphics{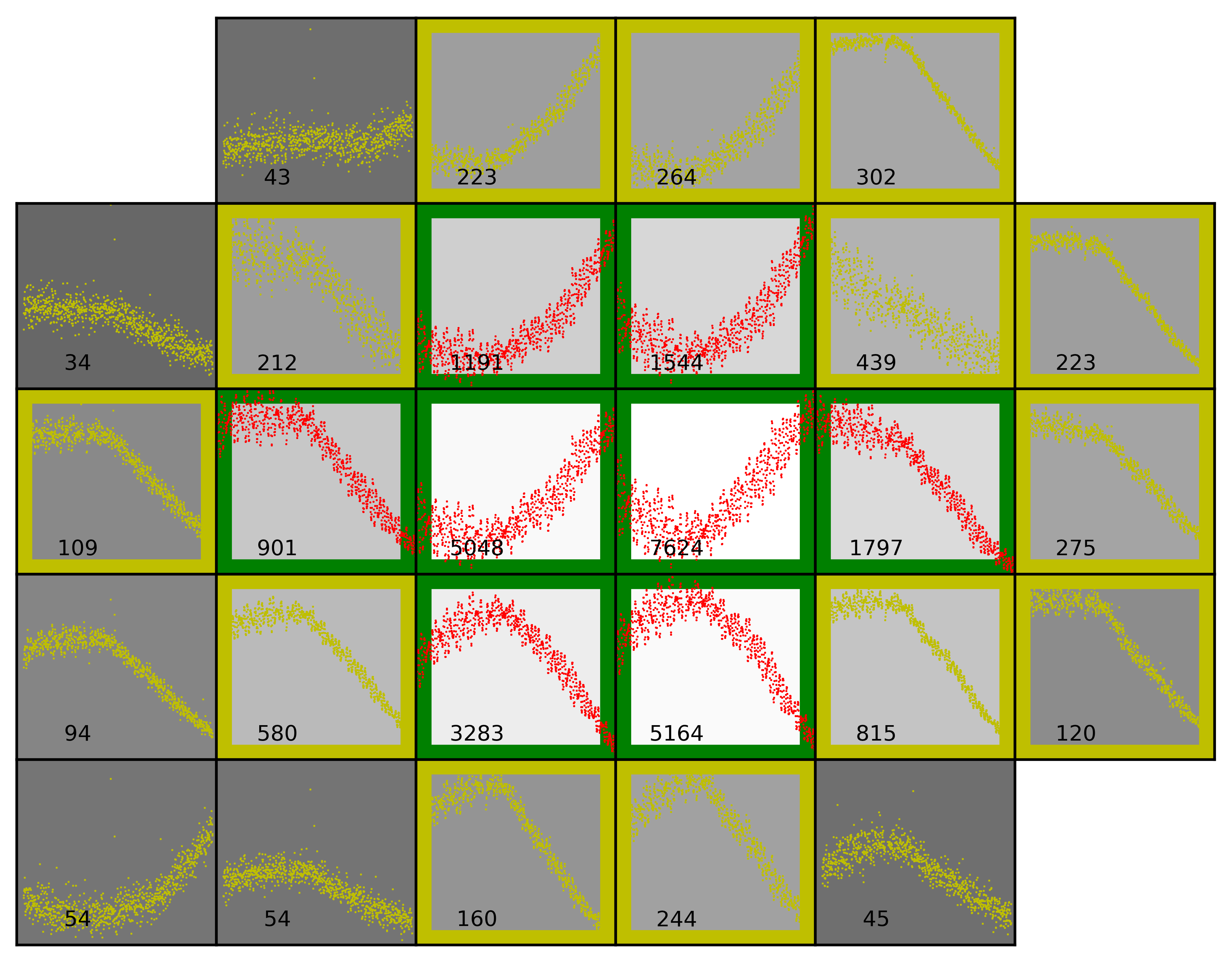}} 
\caption{Pixel mask for the Q5 data of KIC\,10526294. The light curve is plotted for every individual pixel that was downloaded from the spacecraft. The value in every pixel, as well as the background colour, indicates the S/N of the flux in the pixel. The pixels with green borders were used to extract the standard \textit{Kepler} light curves. We have added the yellow pixels (significant signal is present) in our custom mask for the light curve extraction. This results in a light curve with significantly less instrumental effects than the standard extraction.}
\label{pixelmask}
\end{figure}

For the analysis in the forthcoming sections, we used long cadence (LC) data -- 270 cycles of 6.02\,s integration time and 0.52\,s readout time are co-added on board to produce data with a cadence of 29.43\,m -- from Q1 to Q17 (13 May 2009 to 11 May 2013). This is the full \textit{Kepler} dataset of this target. (KIC\,10526294 was not observed during the commissioning Q0.) Light curves were constructed from target pixel files using custom masks (see example on Fig.\,\ref{pixelmask}). Since target pixel files for Q0-Q14 were not corrected in the MAST Archive for the incorrect reporting (see Kepler Data Release 19-20 Notes) of barycentric Julian dates (BJD) in terrestrial dynamic time (TDB), we have carefully corrected the timestamps following the Release Notes -- which we discuss here to provide all necessary information for similar corrections in the future. This was done by adding 66.184\,s to the reported barycentric times for all data points with a cadence interval number (CIN) less than or equal to 57139 in LC (which corresponds to -- in the case of our target -- a reported BKJD\footnote{Barycentric Kepler Julian Date, the time reported by \textit{Kepler} in the target pixel files} $= \mathrm{BJD} - 2454833 = 1276.48872872$), taken during the first month of Q14 at the time of the most recent leap second (UTC 2012-06-30 23:59:60), and by adding 67.184\,s after this cadence. Since data from Q15 onwards had already been reported in the correct BJD\_TDB format, there was no correction necessary beyond Q14. Then quarters were manually cleaned from clear outliers and detrended using a division with a second-order polynomial fit, before finally the counts were converted to ppm, and the quarters were merged into one continuous light curve.

The cleaned and detrended \textit{Kepler} light curve (see Fig.\,\ref{lightcurve}) contains 65236 data points, covers 1459.49\,days ($\sim4$\,years) starting from BJD\,2454964.512312, and has a duty cycle of 91.33\%.

\subsection{Frequency analysis}\label{frequanal}
To extract the frequencies of pulsation modes (and any other possible light variation), we performed a standard iterative prewhitening procedure, whose description is already given by, e.g., \citet{2009A&A...506..111D} and thus not discussed in detail here. This resulted in a list of amplitudes ($A_j$), frequencies ($f_j$), and phases ($\theta_j$). Using these parameters, the light curve can be modelled via $n_f$ frequencies in the well-known form of \[F(t_i)=c+\sum_{j=1}^{n_f}A_j\sin[2\pi(f_j t_i + \theta_j)].\] The prewhitening procedure was stopped when a $p$ value of $p = 0.001$ was reached in hypothesis testing of the significance of the frequecy.

To provide a physically meaningful -- and practically manageable -- subset of frequencies, we limited ourselves to those peaks, which had a S/N of at least 4 in a $1\,\mathrm{d}^{-1}$ window before being removed in the corresponding prewhitening stage. Defining proper significance criteria for high-quality space-based photometry is far from trivial \citep[see, e.g.,][]{2012A&A...542A..55P}, and we cannot exclude the possibility that we have left a few low-amplitude modes out that are intrinsic to the star by adopting this classical approach \citep{1993A&A...271..482B}, but less conservative criteria result in such extensive frequency sets, which -- beyond being unlikely to be 100\% intrinsic to the star -- come with a saturation in frequency of some regions in the \citet{1982ApJ...263..835S} periodogram.

To understand this concern, we recall the \citet{1978Ap&SS..56..285L} criterion, which states that the minimal frequency separation that two close peaks must have to avoid influence on their apparent frequencies in the periodogram is $\sim2.5/T = 0.00171\,\mathrm{d}^{-1}$ (where $T$ is the total timespan of the observations). This means that arising from the combination of the outstanding photometric precision of the \textit{Kepler} light curves and the finite frequency resolution -- also known as the Rayleigh limit -- of the periodogram ($1/T=0.000685\,\mathrm{d}^{-1}$), there might be more peaks appearing within given frequency intervals than what is possible without these peaks influencing each other during their prewhitening \citep[for an illustrative example, see][]{2010_degroote_phd}. To illustrate this for our case, looking at the region between $0.25\,\mathrm{d}^{-1}$ and $2.5\,\mathrm{d}^{-1}$, the number of significant frequencies is 295 using our criterion, but simply using a larger window of $3\,\mathrm{d}^{-1}$ for the significance measurement would result in 1770 significant frequencies instead. While in the default case we have a peak density of $131.11$ peaks per $1\,\mathrm{d}^{-1}$ (an average separation of $0.0076\,\mathrm{d}^{-1}$), in the latter case there would be $786.67$ peaks per $1\,\mathrm{d}^{-1}$ (an average separation of $0.0013\,\mathrm{d}^{-1}$), which means that -- on average -- none of the consecutive peaks would fit the \citet{1978Ap&SS..56..285L} criterion.

By opting for a more conservative approach, we can avoid listing countless frequencies which are influenced by the crowding illustrated above. This means that while the prewhitening process returned 2553 model frequencies, the number of peaks that met our significance criterion is only 346. This set of significant frequencies upon which the rest of the analysis is built upon is listed in Table\,\ref{frequtable}. The variance reduction of the model constructed using this set is 99.68\%, while it brings down the average signal levels from 500.3--53.1--9.1--8.1--7.4 ppm to 19.6--10.7--0.8--0.7--0.6 ppm, measured in $2\,\mathrm{d}^{-1}$ windows centred on 1, 2, 5, 10, and $20\,\mathrm{d}^{-1}$, respectively.

\subsection{The pulsation spectrum}
\begin{figure*}
\resizebox{\hsize}{!}{\includegraphics{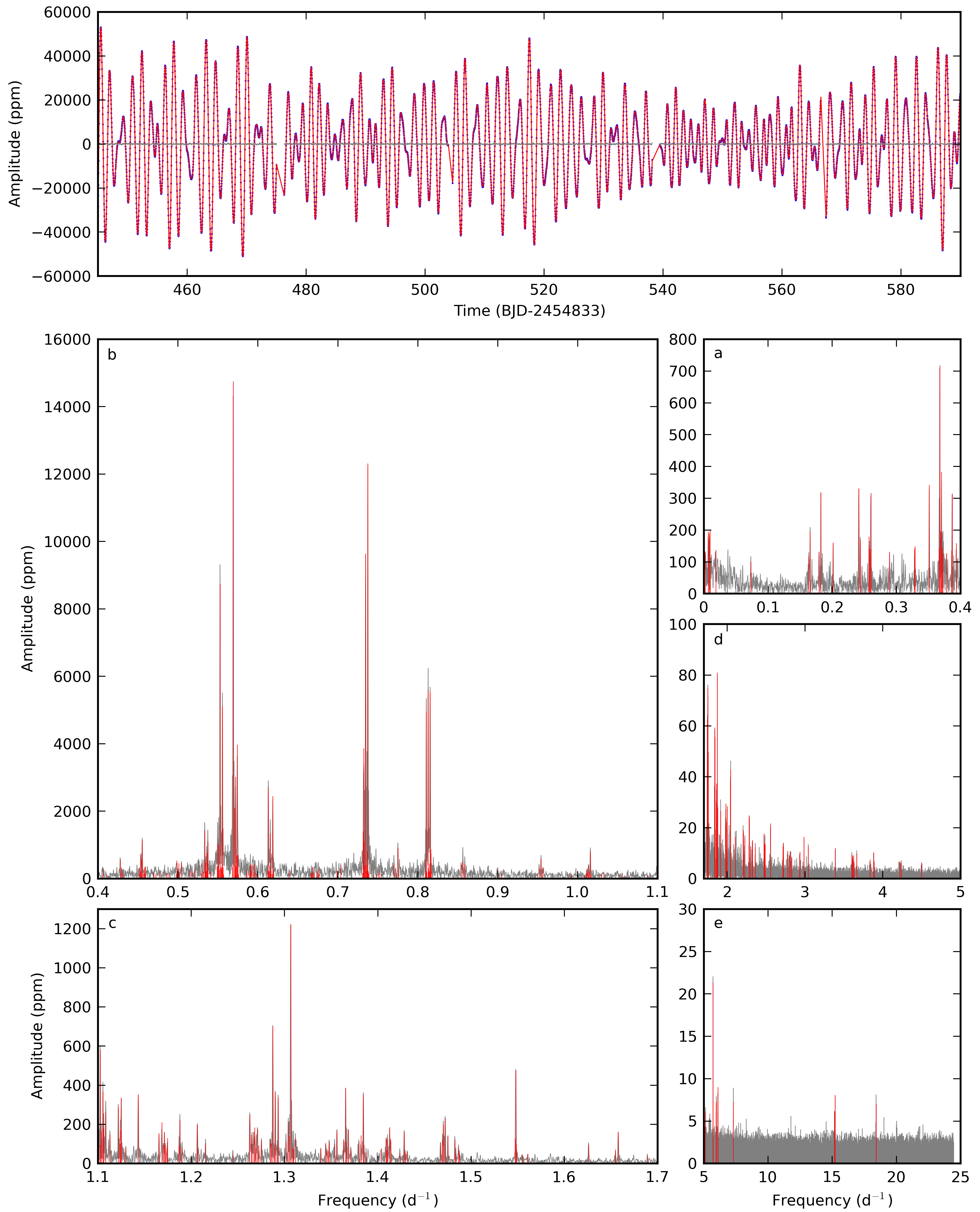}} 
\caption{(\textit{Upper panel}) Zoom-in section showing approximately $1/10\mathrm{th}$ of the reduced \textit{Kepler} light curve (blue dots, brightest at the top) of KIC\,10526294, and the residuals (grey dots) after prewhitening with a model (red solid line) constructed using all model frequencies -- see text for further explanation. (\textit{Lower panels from a to e}) The Scargle periodogram of the full \textit{Kepler} light curve (grey solid line) showing the 346 significant frequencies (red vertical lines). For clarity reasons, the full frequency range is cut into five different parts, and the signal from outside the plotted ranges is prewhitened for each panel. Note the different scales utilised for these panels.}
\label{fourier}
\end{figure*}

The pulsation spectrum (see Fig.\,\ref{fourier}) of KIC\,10526294 is a typical, pure $g$ mode spectrum as expected for SPB stars. The strongest peaks are found between $0.4\,\mathrm{d}^{-1}$ and $0.9\,\mathrm{d}^{-1}$, while there is basically no significant power below $\sim0.15\,\mathrm{d}^{-1}$ and above $\sim2.5\,\mathrm{d}^{-1}$, although the transition from high to low power density is much less pronounced around the high frequency end of the described interval. 

There are three dominant features in the Scargle periodogram. First of all, combination frequencies can be seen above $\sim1.1\,\mathrm{d}^{-1}$ (see Sect.\,\ref{combinations}), which are responsible for the aforementioned smooth transition between high and low power density regions. More importantly, there is a clear series of peaks showing a nearly equidistant spacing in period (see Sect.\,\ref{periodspacing}), and there is also a systematic crowding of pulsation modes around these frequencies, which we interpret as a sign of rotational splitting (see Sect.\,\ref{rotationalsplitting}). 

\subsubsection{Combination frequencies and pressure modes}\label{combinations}
An automated search for linear combinations among the most significant ($\mathrm{S}/\mathrm{N}>8$) subset of frequencies following the approach described by \citet{2012AN....333.1053P} revealed that most of the frequencies situated above $1.1\,\mathrm{d}^{-1}$ can be reproduced as low-order combinations. Out of the 59 frequencies in the subset, all 40 peaks below $1.1\,\mathrm{d}^{-1}$ were found to be independent, while above this limit, 13 peaks were identified as second-order combinations (where $f_i=f_j+f_k$), and two as third-order combinations (where $f_i=2f_j+f_k$ or $f_i=3f_j$). Only the four remaining peaks were not found to be low (second or third) order combinations. Since the ratio of true low-order combination frequencies beyond $1.1\,\mathrm{d}^{-1}$ is very high compared to what is expected from a random set of frequencies, it is more likely that these remaining modes are also actual (higher order) combination frequencies connected to the high amplitude $g$ modes. Lowering the significance threshold to include all significant frequencies and allowing higher order combinations to be found by our automated search, we can indeed match all modes towards higher frequencies as simple combinations, except for the three peaks above $15\,\mathrm{d}^{-1}$.

These three modes could be low amplitude $p$ modes, which would be a significant aid in our seismic modelling (see Sect.\,\ref{seismicmodelling}), but except for their observed frequencies, we can only argue against this possibility. First of all, although these peaks meet our significance criterion for the full data set, when an identical frequency analysis is performed for two halves of the full light curve separately, they can only be found in the list of model frequencies of one half, not both, even without applying any significance criteria. Furthermore, none of these frequencies show rotationally split structures, which would resemble the behaviour we discuss for the $g$ modes in Sect.\,\ref{rotationalsplitting}, although this would be expected for the $\ell\neq0$ $p$ modes, too. Finally, at the cool edge of the SPB instability strip, $p$ modes are not expected to be excited by current excitation calculations. These results imply that only the $g$ modes below $1.1\,\mathrm{d}^{-1}$ need further detailed investigation.

\subsubsection{Period spacing of gravity modes}\label{periodspacing}
\begin{figure}
\resizebox{\hsize}{!}{\includegraphics{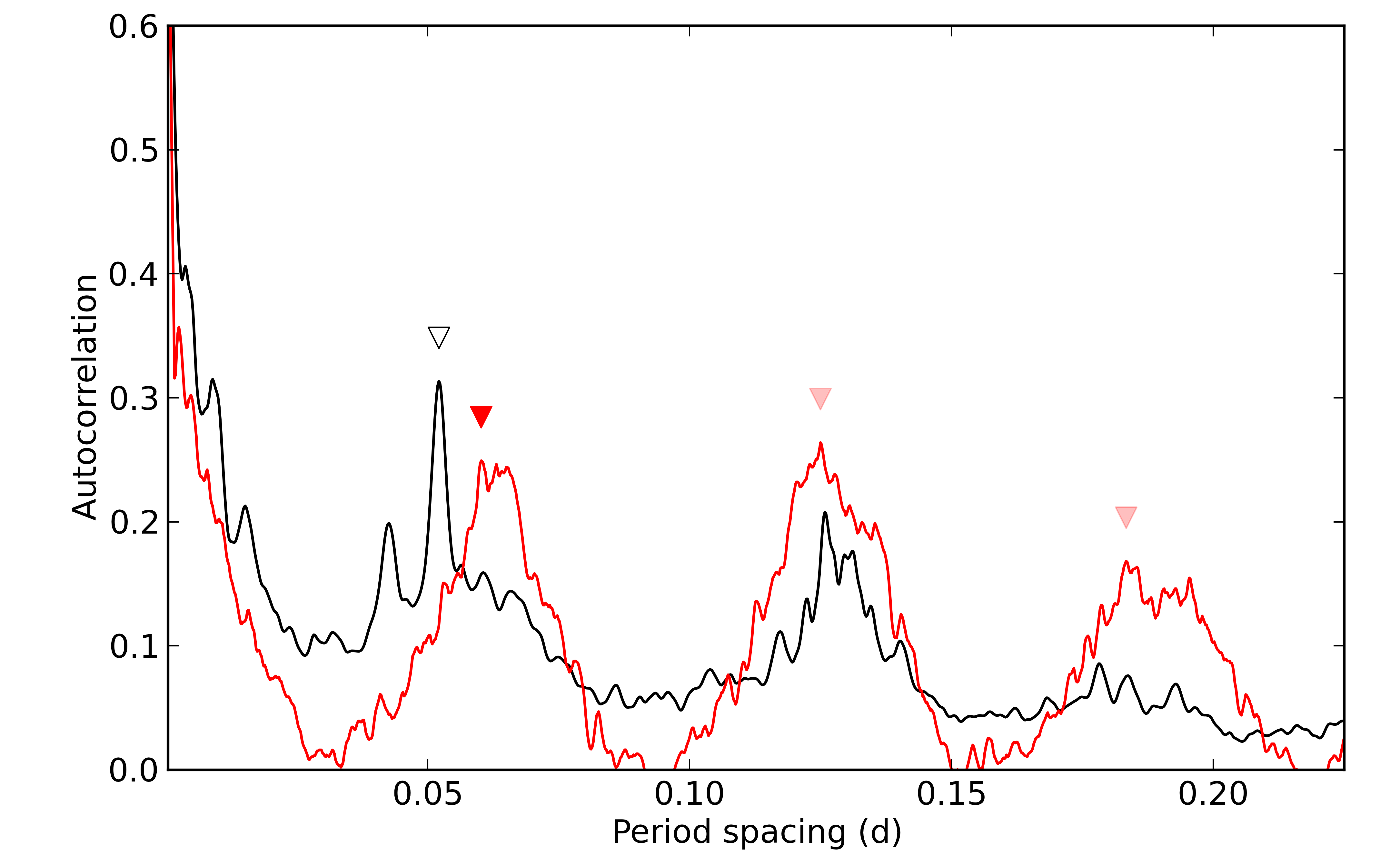}} 
\caption{Autocorrelation function (in period space) of the original Scargle periodogram (black solid line) and of a modified periodogram where the 346 significant frequencies were given an artificial power, while all others were given zero power (red solid line), both calculated from the frequency range of $0.15\,\mathrm{d}^{-1}$ to $1.1\,\mathrm{d}^{-1}$. Candidate spacings are marked with triangles using a colour according to the corresponding function. See text for further explanation.}
\label{autocorrperiod}
\end{figure}

The autocorrelation function of the Scargle periodogram in period space shows a clear structure. We calculated the autocorrelations displayed on Fig.\,\ref{autocorrperiod} in two slightly different ways. After computing the autocorrelation of the original periodogram (transformed into period space), we recalculated the autocorrelation of a modified periodigram, where all the 346 significant peaks were given the same artificial power, and all period values that are more than half the \citet{1978Ap&SS..56..285L} criterion away from the 346 peaks under consideration were placed at value zero. This way we can avoid an autocorrelation function that is dominated by only a few high amplitude peaks, and we can reveal the real structure in the spacing of the significant peaks.

The autocorrelation of the original periodogram peaks at $\delta_0 p = 0.0521\,\mathrm{d}$, but as suggested before, this value is strongly influenced by the two strong peaks at $0.552608\,\mathrm{d}^{-1}$ and $0.569070\,\mathrm{d}^{-1}$, having a separation in period space of $\delta p = 0.0523\,\mathrm{d} \approx \delta_0 p$. It would be better to look at the autocorrelation of the modified periodogram, which shows the following maxima: $\delta_1 p = 0.0602\,\mathrm{d}$, $\delta_2 p = 0.1250\,\mathrm{d}$, and $\delta_3 p = 0.1834\,\mathrm{d}$. Supposing that these are the spacings between consecutive peaks, between every second and every third peak, respectively, in a series of more-or-less equally spaced peaks, then our estimate of the average spacing can be made as $\Delta p \approx [\delta_1 p + (\delta_2 p)/2 + (\delta_3 p)/3]/3 = 0.06128\,\mathrm{d}$ (5295\,s).

\begin{figure}
\resizebox{\hsize}{!}{\includegraphics{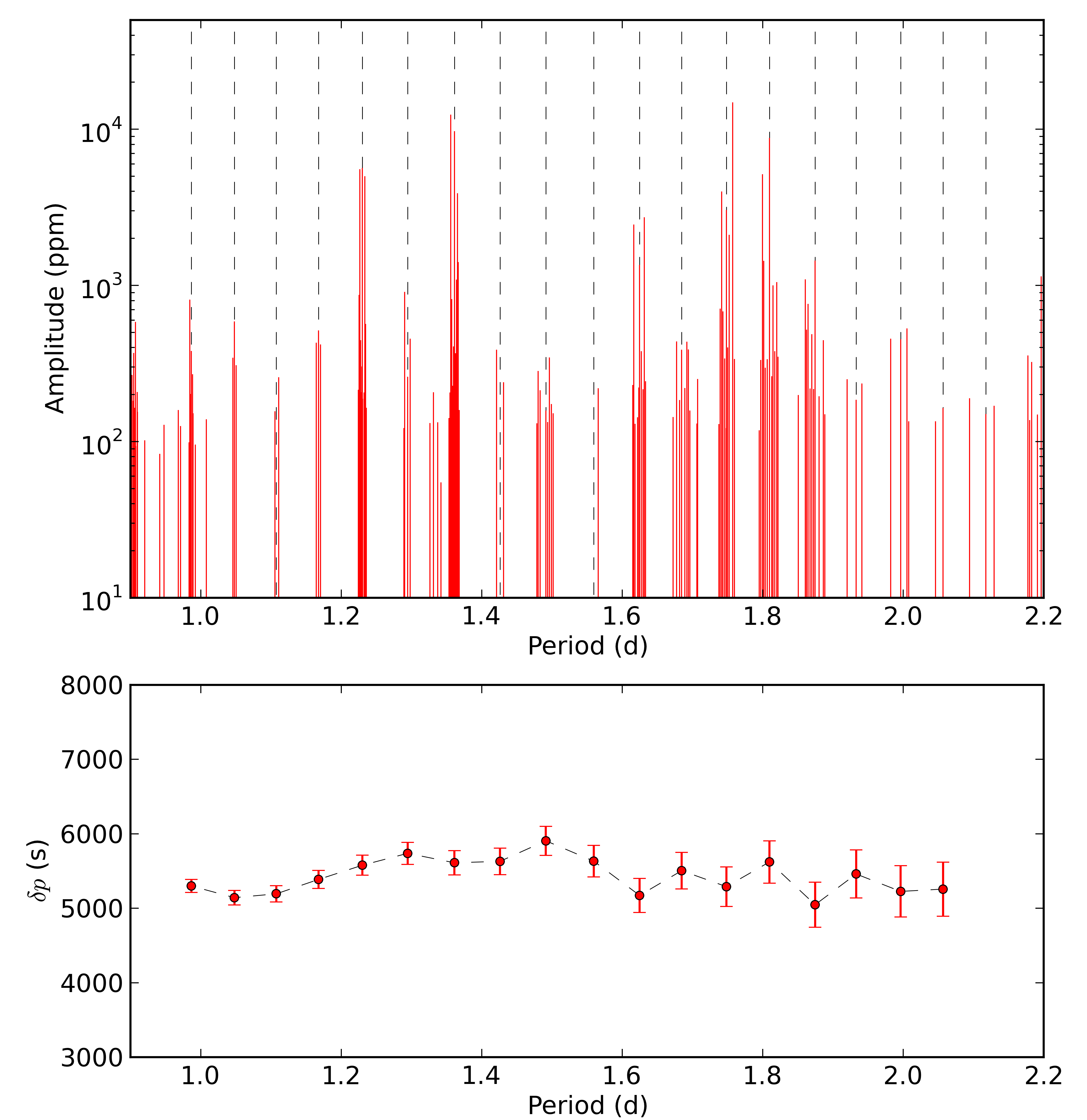}} 
\caption{(\textit{Upper panel}) Period spectrum showing the significant peaks in the periodogram of KIC\,10526294 (red solid lines) with the locations of the members of the series having an almost constant period spacing (grey dashed lines). (\textit{Lower panel}) Individual spacing values between the members of the series (red filled circles).}
\label{periodspacingfig}
\end{figure}

Since not only the average period spacing, but also the deviations from this average carry information about the physical conditions inside the star \citep[see, e.g.,][]{2008MNRAS.386.1487M}, we performed a manual peak selection to reconstruct the full series of nearly equally spaced frequencies. With this method, we found a series of nineteen peaks (which are listed in Table\,\ref{periodspacedpeaks} and marked in Fig.\,\ref{periodspacingfig}) having an average period spacing of $\Delta p = 0.06283\,\mathrm{d}$ (5428\,s). We found small deviations from the exact spacing value (see Fig.\,\ref{periodspacingfig}), with a standard deviation of $\mathrm{std}(\Delta p) = 0.00267\,\mathrm{d}$ (231\,s). We note that although $f_{10}$, $f_{12}$, and $f_{17}$ fail to meet our strict significance criterion by having an S/N of only  3.6, 3.2, and 3.8, respectively, we have included them in our series because they are not only a perfect fit, but are also clearly central peaks of rotationally split triplets where all other components (except for $f_{10+}$ with a S/N of $3.4$) meet the significance criterion (see Sect.\,\ref{rotationalsplitting} and Fig.\,\ref{rotationalsplittingzoominfig}). Furthermore, thanks to the exceptionally high number of consecutive peaks in the series, the uncertainty of the average period spacing is very low. We estimated this value from a Monte Carlo simulation (taking the frequency resolution of the periodogram into account) to be $\epsilon_{\Delta p} = 0.00017\,\mathrm{d}$ (15\,s).

\begin{table}
\caption{Parameters, such as frequencies ($f_j$), periods ($p_j$), and the period spacing (of peaks $f_{j}-f_{j-1}$) values (with their conservative estimated errors based on the Rayleigh limit of $0.000685\,\mathrm{d}^{-1}$) of peaks having a nearly equal spacing in period space in the periodogram of KIC\,10526294.} 
\label{periodspacedpeaks}
\centering
\begin{tabular}{l c c c c}
\hline\hline
ID & $f\,(\mathrm{d}^{-1})$ & $p\,(\mathrm{d})$ & $\delta p\,(\mathrm{d})$ & $\epsilon_{\delta p}\,(\mathrm{d})$\\
\hline
$f_{1}$ &$0.472220 $&$ 2.117657 $&& \\
$f_{2}$ &$0.486192 $&$ 2.056801 $&$ 0.060856$&$ 0.004207 $\\
$f_{3}$ &$0.500926 $&$ 1.996303 $&$ 0.060498$&$ 0.003982 $\\	
$f_{4}$ &$0.517303 $&$ 1.933103 $&$ 0.063200$&$ 0.003737 $\\
$f_{5}$ &$0.533426 $&$ 1.874674 $&$ 0.058429$&$ 0.003514 $\\
$f_{6}$ &$0.552608 $&$ 1.809601 $&$ 0.065073$&$ 0.003290 $\\
$f_{7}$ &$0.571964 $&$ 1.748362 $&$ 0.061239$&$ 0.003070 $\\
$f_{8}$ &$0.593598 $&$ 1.684642 $&$ 0.063720$&$ 0.002851 $\\
$f_{9}$ &$0.615472 $&$ 1.624769 $&$ 0.059873$&$ 0.002660 $\\
$f_{10}$&\textit{0.641202}&$ 1.559571 $&$ 0.065198$&$ 0.002458 $\\
$f_{11}$&$0.670600 $&$ 1.491202 $&$ 0.068369$&$ 0.002257 $\\
$f_{12}$&\textit{0.701246}&$ 1.426033 $&$ 0.065169$&$ 0.002062 $\\	
$f_{13}$&$0.734708 $&$ 1.361085 $&$ 0.064948$&$ 0.001881 $\\
$f_{14}$&$0.772399 $&$ 1.294668 $&$ 0.066417$&$ 0.001715 $\\
$f_{15}$&$0.812940 $&$ 1.230103 $&$ 0.064565$&$ 0.001552 $\\
$f_{16}$&$0.856351 $&$ 1.167745 $&$ 0.062358$&$ 0.001397 $\\
$f_{17}$&\textit{0.902834}&$ 1.107623 $&$ 0.060122$&$ 0.001254 $\\
$f_{18}$&$0.954107 $&$ 1.048100 $&$ 0.059523$&$ 0.001126 $\\
$f_{19}$&$1.013415 $&$ 0.986763 $&$ 0.061338$&$ 0.001004 $\\
\hline
\end{tabular}
\tablefoot{Although frequencies listed in italic have slightly lower significance than what is required by our strict criterion, we have included them in the series for reasons discussed in the text.}
\end{table}

\subsubsection{Rotational splitting of gravity modes}\label{rotationalsplitting}
\begin{figure}
\resizebox{\hsize}{!}{\includegraphics{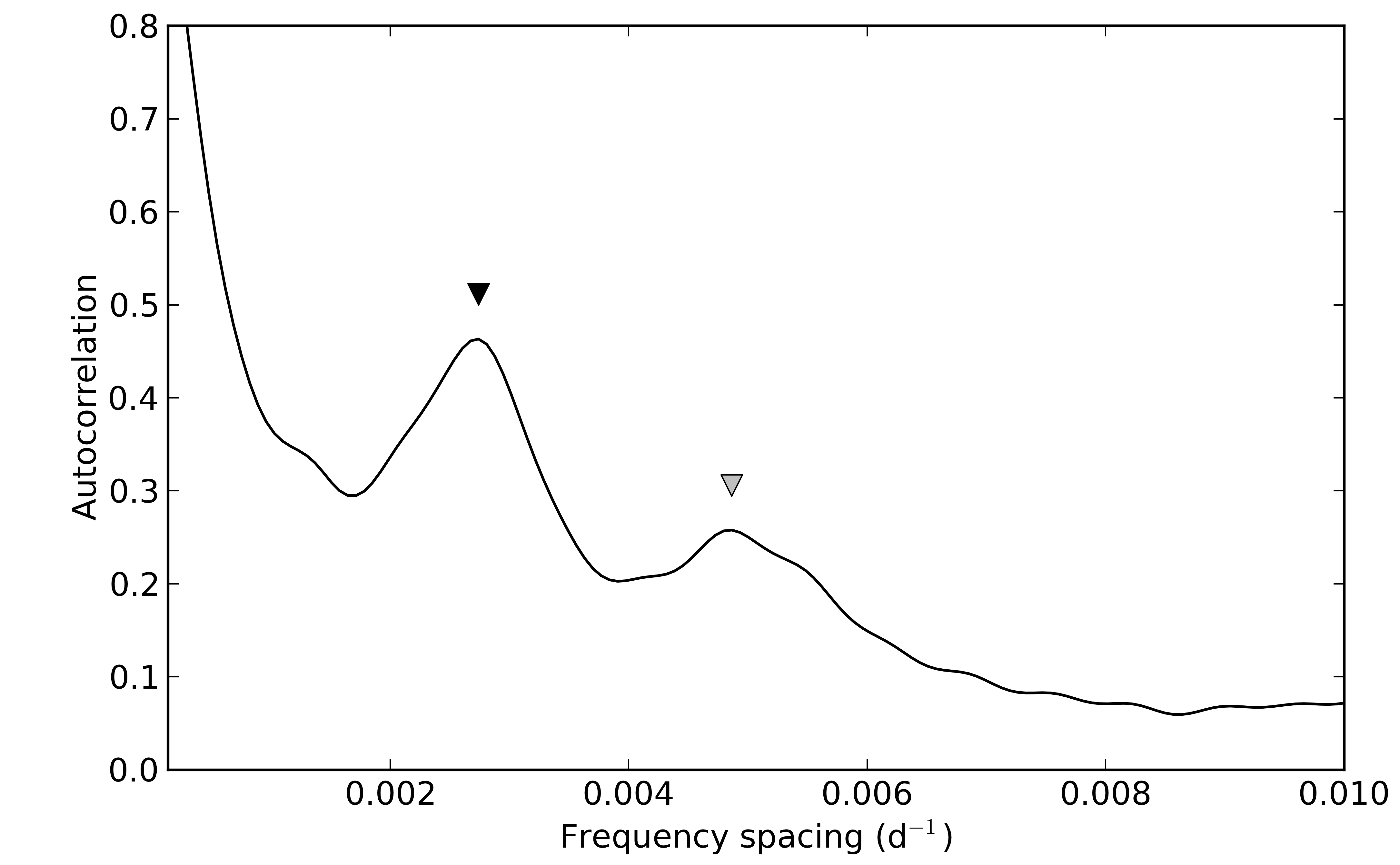}} 
\caption{Autocorrelation function (in frequency space) of the original Scargle periodogram between $0.15\,\mathrm{d}^{-1}$ and $1.1\,\mathrm{d}^{-1}$. The two candidate splitting values are marked with triangles.}
\label{autocorrfourier}
\end{figure}

The autocorrelation function of the peridogram in frequency space in Fig.\,\ref{autocorrfourier} shows two clear peaks at low frequencies: $\delta_1 f = 0.0027\,\mathrm{d}^{-1}$ and $\delta_2 f = 0.0049\,\mathrm{d}^{-1}$. Supposing that these are the spacings between a central ($m=0$) peak and a rotationally split side peak ($0 < |m| \le \ell$) and between two side peaks (of $+m$ and $-m$), respectively, we can estimate the average rotational splitting as $\Delta f \approx [\delta_1 f + (\delta_2 f)/2]/2 = 0.00259\,\mathrm{d}^{-1}$.

\begin{figure}
\resizebox{\hsize}{!}{\includegraphics{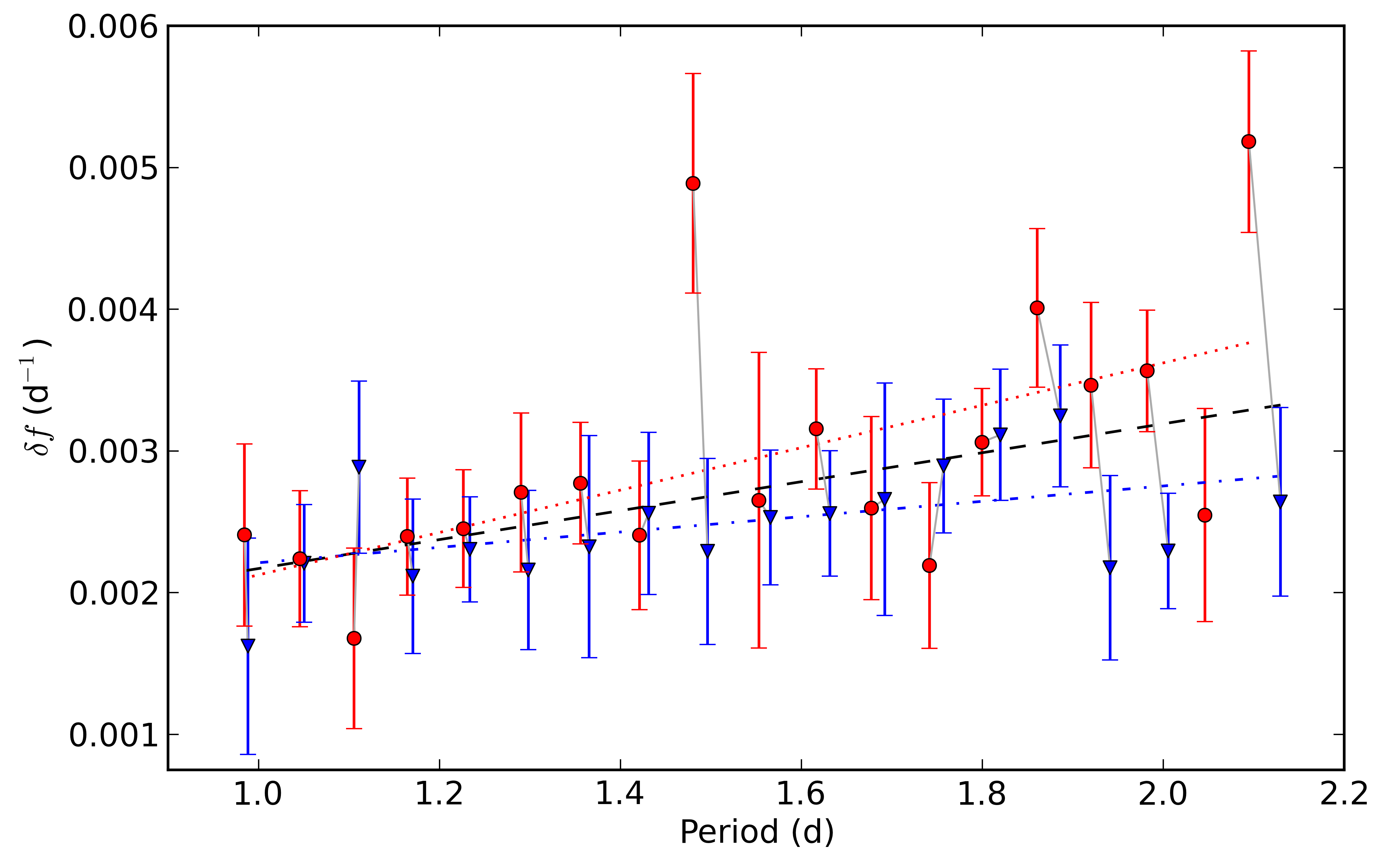}} 
\caption{The rotational splitting values (red filled circles for $f_{j+}$ and blue filled triangles for $f_{j-}$) from the periodogram of KIC\,10526294. Splittings connected to the same central ($m=0$, $f_{j}$) peak are connected with a solid light grey line. Linear fits are plotted for the $m=-1$ peaks (dot-dashed blue line), the $m=+1$ peaks (dotted red line), and the combined dataset (dashed black line).}
\label{rotationalsplittingfig}
\end{figure}

Because different modes penetrate different depths in the star, it is worth looking at not only the average splitting, but also at individual splitting values, in the hope of unraveling changes in the rotation rate in function of the depth. We found that all nineteen peaks in Table\,\ref{periodspacedpeaks} show signatures that can be interpreted as rotational splitting. The average splitting is $\Delta f = 0.00266\,\mathrm{d}^{-1}$, and the individual splittings (see Table\,\ref{rotationallysplitpeaks} and Fig.\,\ref{rotationalsplittingfig}) show a small but clearly increasing trend towards higher period values. This implies a non-rigid internal rotation profile for the star. Nevertheless, we estimated the rotation period of the star averaged throughout its interior from the average of the frequency splitting by substituting $\ell=1$, $m=1$ into \[\Delta f = m\beta_{nl}\Omega \simeq m\left(1-\frac{1}{\ell\left(\ell+1\right)}\right)\Omega=0.5\Omega,\] which gives $P=1/\Omega\approx188\,\mathrm{d}$. For solid body rotation, this would correspond to the rotation period of the stellar surface, but for non-rigid rotation, this is an average value dominated by the internal layers because that is where the observed $g$ modes have their largest contribution to the splitting.

\begin{table}
\caption{Parameters, such as frequencies ($f_j$), periods ($p_j$), the splitting value (of side peaks $f_{j+}$ and $f_{j-}$ measured from the central $m=0$ peak $f_{j}$), and the average splitting of the rotationally split triplets in the periodogram of KIC\,10526294.} 
\label{rotationallysplitpeaks}
\centering
\begin{tabular}{l c c c c}
\hline\hline
ID & $f\,(\mathrm{d}^{-1})$ & $p\,(\mathrm{d})$ & $\delta f\,(\mathrm{d}^{-1})$ & $\langle\delta f\rangle\,(\mathrm{d}^{-1})$\\
\hline
$f_{1-}$ &$0.469579$&$2.129567$& $0.002641$&\\
$f_{1 }$ &$0.472220$&$2.117657$&&$0.003912$\\
$f_{1+}$ &$0.477403$&$2.094666$& $0.005183$&\\
$f_{2-}$ &          &          &           &\\
$f_{2 }$ &$0.486192$&$2.056801$&&          \\
$f_{2+}$ &$0.488740$&$2.046078$& $0.002548$&\\
$f_{3-}$ &$0.498632$&$2.005487$& $0.002294$&\\
$f_{3 }$ &$0.500926$&$1.996303$&&$0.002930$\\
$f_{3+}$ &$0.504491$&$1.982196$& $0.003565$&\\
$f_{4-}$ &$0.515126$&$1.941273$& $0.002177$&\\
$f_{4 }$ &$0.517303$&$1.933103$&&$0.002821$\\
$f_{4+}$ &$0.520767$&$1.920245$& $0.003464$&\\
$f_{5-}$ &$0.530179$&$1.886155$& $0.003247$&\\
$f_{5 }$ &$0.533426$&$1.874674$&&$0.003628$\\
$f_{5+}$ &$0.537435$&$1.860690$& $0.004009$&\\
$f_{6-}$ &$0.549494$&$1.819856$& $0.003114$&\\
$f_{6 }$ &$0.552608$&$1.809601$&&$0.003088$\\
$f_{6+}$ &$0.555670$&$1.799629$& $0.003062$&\\
$f_{7-}$ &$0.569070$&$1.757253$& $0.002894$&\\
$f_{7 }$ &$0.571964$&$1.748362$&&$0.002543$\\
$f_{7+}$ &$0.574156$&$1.741687$& $0.002192$&\\
$f_{8-}$ &$0.590939$&$1.692222$& $0.002659$&\\
$f_{8 }$ &$0.593598$&$1.684642$&&$0.002628$\\
$f_{8+}$ &$0.596196$&$1.677301$& $0.002598$&\\
$f_{9-}$ &$0.612913$&$1.631553$& $0.002559$&\\
$f_{9 }$ &$0.615472$&$1.624769$&&$0.002857$\\
$f_{9+}$ &$0.618628$&$1.616480$& $0.003156$&\\
$f_{10-}$&$0.638671$&$1.565751$& $0.002531$&\\
$f_{10 }$&\textit{0.641202}&$1.559571$&&$0.002592$\\
$f_{10+}$&\textit{0.643854}&$1.553147$& $0.002652$&\\
$f_{11-}$&$0.668309$&$1.496314$& $0.002291$&\\
$f_{11 }$&$0.670600$&$1.491202$&&$0.003590$\\
$f_{11+}$&$0.675489$&$1.480409$& $0.004889$&\\
$f_{12-}$&$0.698686$&$1.431258$& $0.002560$&\\
$f_{12 }$&\textit{0.701246}&$1.426033$&&$0.002483$\\
$f_{12+}$&$0.703651$&$1.421159$& $0.002405$&\\
$f_{13-}$&$0.732383$&$1.365406$& $0.002325$&\\
$f_{13 }$&$0.734708$&$1.361085$&&$0.002549$\\
$f_{13+}$&$0.737481$&$1.355967$& $0.002773$&\\
$f_{14-}$&$0.770238$&$1.298300$& $0.002161$&\\
$f_{14 }$&$0.772399$&$1.294668$&&$0.002435$\\
$f_{14+}$&$0.775107$&$1.290144$& $0.002708$&\\
$f_{15-}$&$0.810634$&$1.233602$& $0.002306$&\\
$f_{15 }$&$0.812940$&$1.230103$&&$0.002379$\\
$f_{15+}$&$0.815392$&$1.226404$& $0.002452$&\\
$f_{16-}$&$0.854235$&$1.170638$& $0.002116$&\\
$f_{16 }$&$0.856351$&$1.167745$&&$0.002256$\\
$f_{16+}$&$0.858747$&$1.164487$& $0.002396$&\\
$f_{17-}$&$0.899948$&$1.111175$& $0.002886$&\\
$f_{17 }$&\textit{0.902834}&$1.107623$&&$0.002282$\\
$f_{17+}$&$0.904512$&$1.105569$& $0.001678$&\\
$f_{18-}$&$0.951900$&$1.050531$& $0.002207$&\\
$f_{18 }$&$0.954107$&$1.048100$&&$0.002223$\\
$f_{18+}$&$0.956347$&$1.045646$& $0.002240$&\\
$f_{19-}$&$1.011792$&$0.988345$& $0.001623$&\\
$f_{19 }$&$1.013415$&$0.986763$&&$0.002015$\\
$f_{19+}$&$1.015822$&$0.984424$& $0.002407$&\\
\hline
\end{tabular}
\end{table}

\begin{figure*}
\resizebox{\hsize}{!}{\includegraphics{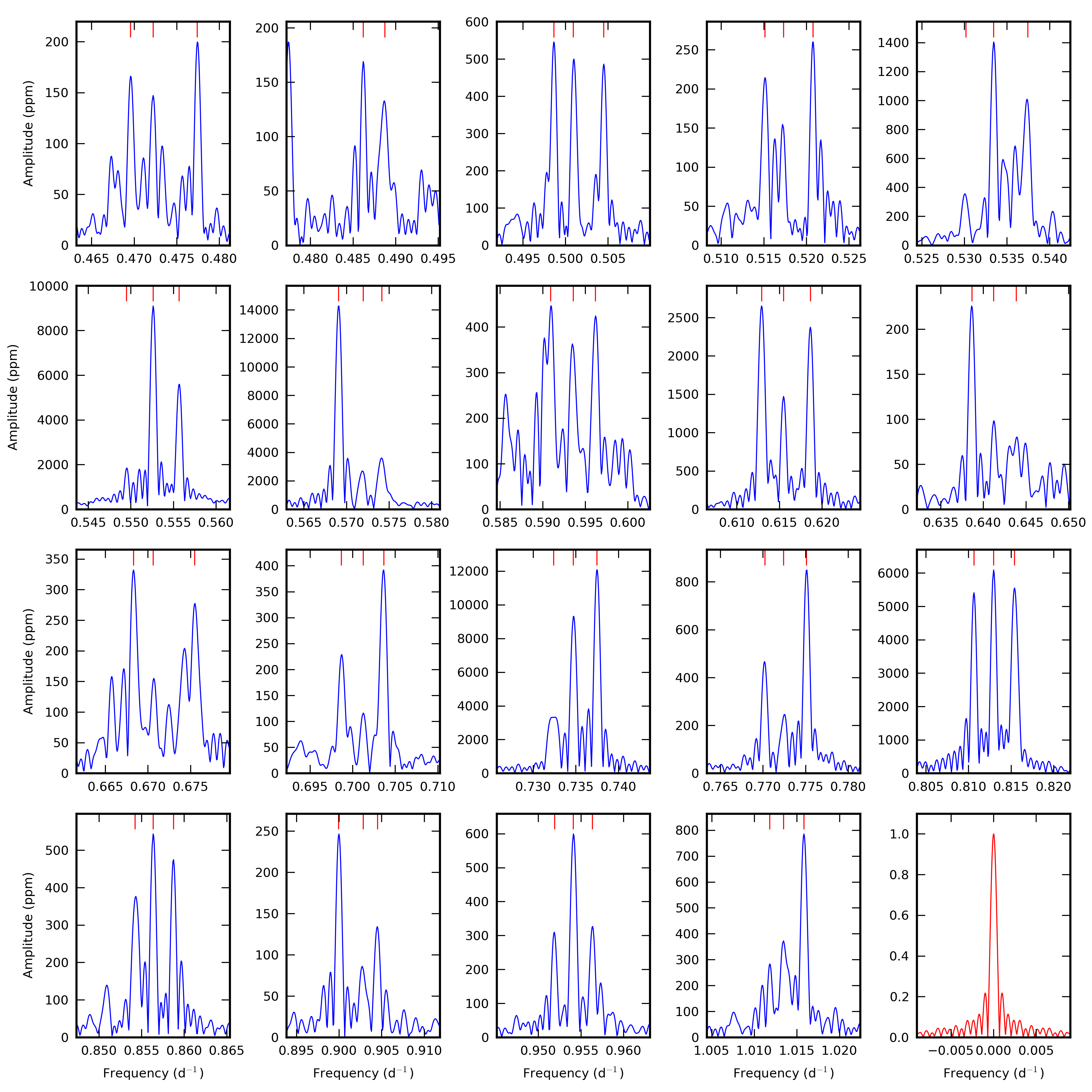}} 
\caption{All rotationally split modes in the Scargle periodogram (blue solid line) of KIC\,10526294, ranging from barely detectable and strongly asymmetric to strong and very clear cases, covering a broad frequency range. Each subplot is centred on the central ($m=0$, $f_{j}$) peak, has the same horizontal scale, and all signal from outside the plotted intervals is prewhitened. Frequencies identified as members of rotationally split triplets (from Table\,\ref{rotationallysplitpeaks}) are marked with red markers on the top of each subplot. The window function of the data is plotted in the lower right panel for comparison (using a red solid line), scaled to have a maximum amplitude of unity.}
\label{rotationalsplittingzoominfig}
\end{figure*}

Another noteworthy trend is the observed asymmetry of the rotational splittings of the $f_{j+}$ and $f_{j-}$ modes. This detection must be taken with caution, since most asymmetries (the difference between the splitting observed for the $f_{j+}$ and $f_{j-}$ modes) are close to the Rayleigh limit, while even the average splitting is only $\sim1.5$ times greater than the \citet{1978Ap&SS..56..285L} criterion. To give a more robust error estimate for the frequencies of the individual components of our triplets, and thus the individual rotational splittings, we did the following. We first calculated the full width at half maximum (FWHM) of the central peak of the window function ($0.000832\,\mathrm{d}^{-1}$ -- slightly above the Rayleigh limit), then calculated the ratio of the area of a box having the height of the central peak and the width of $3\times\mathrm{FWHM}$ and the integral of the window function within the limits of this box to be $r_0\approx2.518$. For a real peak in the observed periodogram, which is situated far enough from other peaks to be unaffected by them, one would expect the same ratio as the result of a similar procedure ($r_i \approx r_0$). For a peak that is surrounded by other very closely spaced peaks, one would expect a lower ratio ($r_i < r_0$), since the box around the peak in question will be filled with signal coming from other peaks, too. We used $(r_0/r_i)^2$ as a multiplier factor to the original FWHM to estimate the uncertainty in the frequencies originating from the crowding around certain peaks. The error bars shown in Fig.\,\ref{rotationalsplittingfig} come from these individual error estimates.

To see if the order of prewhitening had any significant influence on the closely spaced frequencies, we carried out the following check. We prewhitened the signal (using all model frequencies from the original prewhitening) from outside a narrow range around a selected triplet, then did a new iterative prewhitening to find the frequencies within this narrow range. We did this for four selected cases (ranging from the best to the worst case scenarios -- see Fig.\,\ref{rotationalsplittingzoominfig}), then compared the frequencies from the original prewhitening and the new narrow prewhitening procedures. We found that even in the worst case, the original and new frequency values were matching each other with an order-of-magnitude better precision than the Rayleigh frequency, and for the stronger triplets, the match was even another order of magnitude better.

We also performed a nonlinear least square fit using the 56 modes (from Table\,\ref{rotationallysplitpeaks}), which were identified as members of rotationally split triplets, to see if the refined values deviate significantly from the original parameters provided by the automated iterative prewhitening procedure (as discussed in Sect.\,\ref{frequanal}). We found that the average difference between the original and fitted frequencies was $4.8\%$ of the Rayleigh limit (with a median of $2.6\%$, and a maximum of $17\%$). This means that all tested methods deliver results that are identical to each other within $\sim10\%$ of the Rayleigh limit (thus resulting in an actual frequency precision of at least, but in most cases even better than $\sim0.00007\,\mathrm{d}^{-1}$), and makes us conclude that the measured splitting values and asymmetries are independent of the frequency analysis method.

We would like to point out that, even after four years of data gathering, the periodograms are influenced by the window function in the sense that the frequency peaks are not delta functions. The observed power spectrum is rather dense, especially near the rotationally split dipole modes owing to the slow rotation of the star. In these regions it is possible that peaks of $\ell > 1$ pulsation modes with amplitudes comparable to the underlying complicated noise spectrum superimposed on the noise itself are also observed. Since $\ell > 1$ $g$-modes have period spacings smaller than the $\ell=1$ modes, this is not expected to happen in all the regions, but it can occur in those where the two series overlap closely enough. This can explain why we detect additional frequencies around or between the three rotationally split components of the dipole modes in some of the frequency regions but not in others (see Fig.\,\ref{periodspacingfig}).

Another remark on the frequency spectrum is also appropriate. The orbital period of the \textit{Kepler} satellite is $372.5\,\mathrm{d}$, which translates into an orbital frequency of $1/372.5\,\mathrm{d} = 0.002685\,\mathrm{d}^{-1}$. The dipole mode splittings we detected range from $0.001678\,\mathrm{d}^{-1}$ to $0.005183\,\mathrm{d}^{-1}$. Given that we worked in the barycentric time frame and that we see no orbital frequency effects in the previously analysed B-type stars that we processed in the same way with the same frequency analysis methods \citep[see, e.g.,][]{2013A&A...553A.127P}, we are confident that the correspondence between the average splitting and the inverse of the \textit{Kepler} orbital period within the frequency uncertainty is just a coincidence, and we excluded any effect of the satellite orbit on the detected triplets.

Finally, we note that we cannot put constraints on the inclination angle of the star because we are dealing with heat-driven modes whose triplet components are not excited to the same amplitude as is the case for solar-like pulsators \citep[e.g.,][]{2013ApJ...766..101C}. This is clearly illustrated by the irregular amplitude structure of the rotationally split components in Fig.\,\ref{rotationalsplittingzoominfig}, where some of the very clean triplets have a dominant central component, while others have a weaker central than wings component. The measured overall isotropic Gaussian spectral line profile broadening of $18\pm4\,\mathrm{km\,s}^{-1}$ (Table\,\ref{fundparams}) is the combined effect of the rotational, turbulent, and pulsational broadening, the last being the collective effect from all the triplet components. For the modes we are dealing with here, it concerns dominantly tangential velocities, while the radial velocity components are typically only 1/70 of the tangential components for the measured frequencies and mass and radius  we deduce in the following section. The rotational line broadening stems from the equatorial \textit{surface} velocity, which may be quite different from the average rotational velocity deduced from the triplets because they probe the internal rotation of the star. (If the star were a rigid rotator, then the rotation velocity at the equator of the best models to be derived in the following section would be $\simeq 1\,\mathrm{km\,s}^{-1}$ such that the dominant line broadening likely originates in the pulsations \citep[cf.\ the case of the SPB star HD\,181558;][]{2005A&A...432.1013D}.) We do not have any means of disentangling the individual surface velocity broadening contributions. On the other hand, the linear response of the pulsation frequencies in the light curve leads us to deduce that the majority of the individual surface velocity components do not exceed the local sound speed in the stellar atmosphere.


\section{Seismic modelling}\label{seismicmodelling}
We performed forward modelling \citep[see, e.g.,][]{2010aste.book.....A} and compared the set of observed $g$ modes to theoretical pulsation modes calculated using the adiabatic frame of the GYRE stellar oscillation code \citep{2013MNRAS.435.3406T} for a large number of non-rotating stellar equilibrium models along evolutionary tracks passing through the spectroscopic error box ($T_\mathrm{eff}$, $\log g$). The models were calculated with the MESA stellar structure and evolution code \citep{2011ApJS..192....3P,2013ApJS..208....4P}. A first set of models is based on the initial chemical composition of ($X, Y, Z$) = ($0.710, 0.276, 0.014$) as derived by \citet{2012A&A...539A.143N,2013EAS....63...13P}. The OPAL Types 1 and 2 tables \citep{1996ApJ...464..943I} are calculated with the elemental abundance mixture given by \citet[]{2012A&A...539A.143N} and \citet[]{2013EAS....63...13P}. Additional sets of models for various initial metal abundances $Z$ other than 0.014 were also computed, keeping the same metal mixture. The convective core overshooting is described using the exponentially decaying prescription of \citet{2000A&A...360..952H}, and the Ledoux criterion is used to distinguish between convective and radiative regions. We used the semi-convective prescription of \citet{1983A&A...126..207L} with $\alpha_{sc}=10^{-2}$, which mimics instantaneous mixing. We also explicitly checked and ensured that the location of the convective core boundary is appropriately calculated, cf.\ \citet{2014arXiv1405.0128G} for a recent discussion on this matter.

\subsection{The grid}
We calculated a set of evolutionary tracks and the oscillation properties of the models along these tracks with a high temporal and spatial (i.e., along the radial direction inside the star) resolution. The four free parameters of the models are the mass, the initial metallicity, the central hydrogen fraction, and the core overshoot value. We experimented extensively on the needed resolution within this four-dimensional grid in order to obtain sufficiently stringent constraints compared to the measured seismic properties of KIC\,10526294. In this way, we ended up with a high-resolution grid containing approximately 330\,000 models -- see Table\,\ref{gridparams} for further details. We set the range for initial metallicity keeping the cosmic abundance standard for B stars in mind \citep{2012A&A...539A.143N}, rather than the poorly constrained $Z$ of the target. We computed the eigenfrequencies of the $\ell=1$ modes, since the observed triplet structures imply that we are dealing with dipole modes. We calculated two different reduced $\chi^2$ values for each model and considered their average as a goodness-of-fit function. The first $\chi^2_{spacings}$ was calculated from the observed $\delta p_i$ values and the theoretically predicted $\delta p(p)$ function interpolated (linear) onto the set of the observed $p_i$ periods (which have corresponding $\delta p_i$ values), while the other $\chi^2_{periods}$ was calculated relying only on the observed and theoretically predicted dipole mode periods, taking the corresponding error estimates into account in both cases, following the formulae below: \[\chi^2_{spacings} = \frac{1}{18-4}\sum_{i=1}^{18} \frac{\left[\delta p_{i,obs} - \delta p_{theo}\left(p_{i,obs}\right)\right]^2}{\epsilon_{\delta p_i}^2}\] \[\chi^2_{periods} = \frac{1}{19-4}\sum_{i=1}^{19} \frac{\left(p_{i,obs} - p_{i,theo}\right)^2}{\epsilon_{p_i}^2}\] \[\chi^2=\frac{\chi^2_{function}+\chi^2_{periods}}{2}.\] 

\begin{table}
\caption{Parameters of the MESA/GYRE grid. For each parameter, the minimum and maximum values are given along with the corresponding resolution.}
\label{gridparams}
\centering
\renewcommand{\arraystretch}{1.1}
\begin{tabular}{l l l l}
\hline\hline
Parameter&min.&max.&$\delta$\\
\hline
Mass ($\mathcal{M}_\odot$)&$3.00$&$3.40$&$0.05$\\
Core overshoot value&$0.000$&$0.030$&$0.003$\\
Initial metallicity&$0.010$&$0.020$&$0.001$\\
Central hydrogen fraction&$0.400$&$0.700$&$0.001$\\
\hline
\end{tabular}
\end{table}

In the ideal case of a grid with infinite resolution, and given the assumption that a perfect fit exists somewhere in the $N=4$ dimensional parameter space, these two $\chi^2$ tests result in the very same minimum, thereby selecting this best fit model point. As the resolution of the grid degrades, one quickly loses the ability to fit the individual frequencies perfectly, but the fit to the $\delta p(p)$ function -- or the fit of the average period spacing -- will stay quite good for a somewhat coarser grid (depending mostly on the age of the actual best fit model) as we move farther away from the location of the best parameter combination in the $N=4$ dimensional parameter space.

\subsection{Results and discussion}
\begin{figure*}
\resizebox{\hsize}{!}{\includegraphics{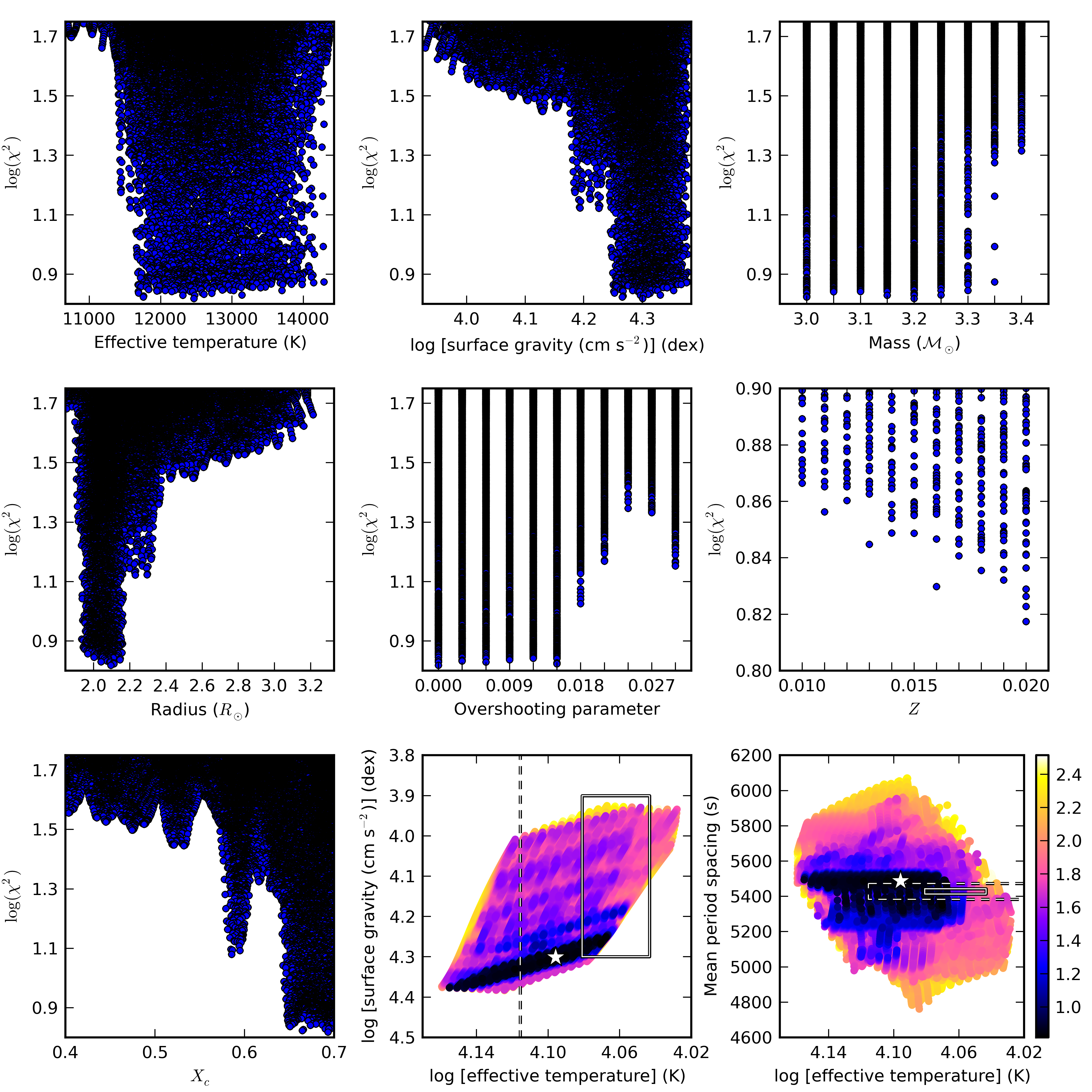}} 
\caption{Visualisation of the resulting $\chi^2$ space from the multi-dimensional grid search using the high-resolution MESA/GYRE grid. The first seven panels (from left to right and top to bottom) show the dependence of the $\chi^2$ value on effective temperature, surface gravity, mass, radius, overshooting, metallicity, and central hydrogen content of the corresponding model. The last two panels show the $\chi^2$ distribution of the model points over the Kiel diagram, and over a $\log T_\mathrm{eff}$ versus measured average period spacing diagram.  The colour of each symbol represents the logarithm of the corresponding $\chi^2$ value. The $1$-$\sigma$ and $3$-$\sigma$ error boxes of KIC\,10526294 in $\log T_\mathrm{eff}$, $\log g$, and mean period spacing are plotted using solid and dashed lines, respectively. The position of the best model is marked with a white asterisk. For further details, see explanation in the text.}
\label{MESAchi2}
\end{figure*}

\begin{figure}
\resizebox{\hsize}{!}{\includegraphics{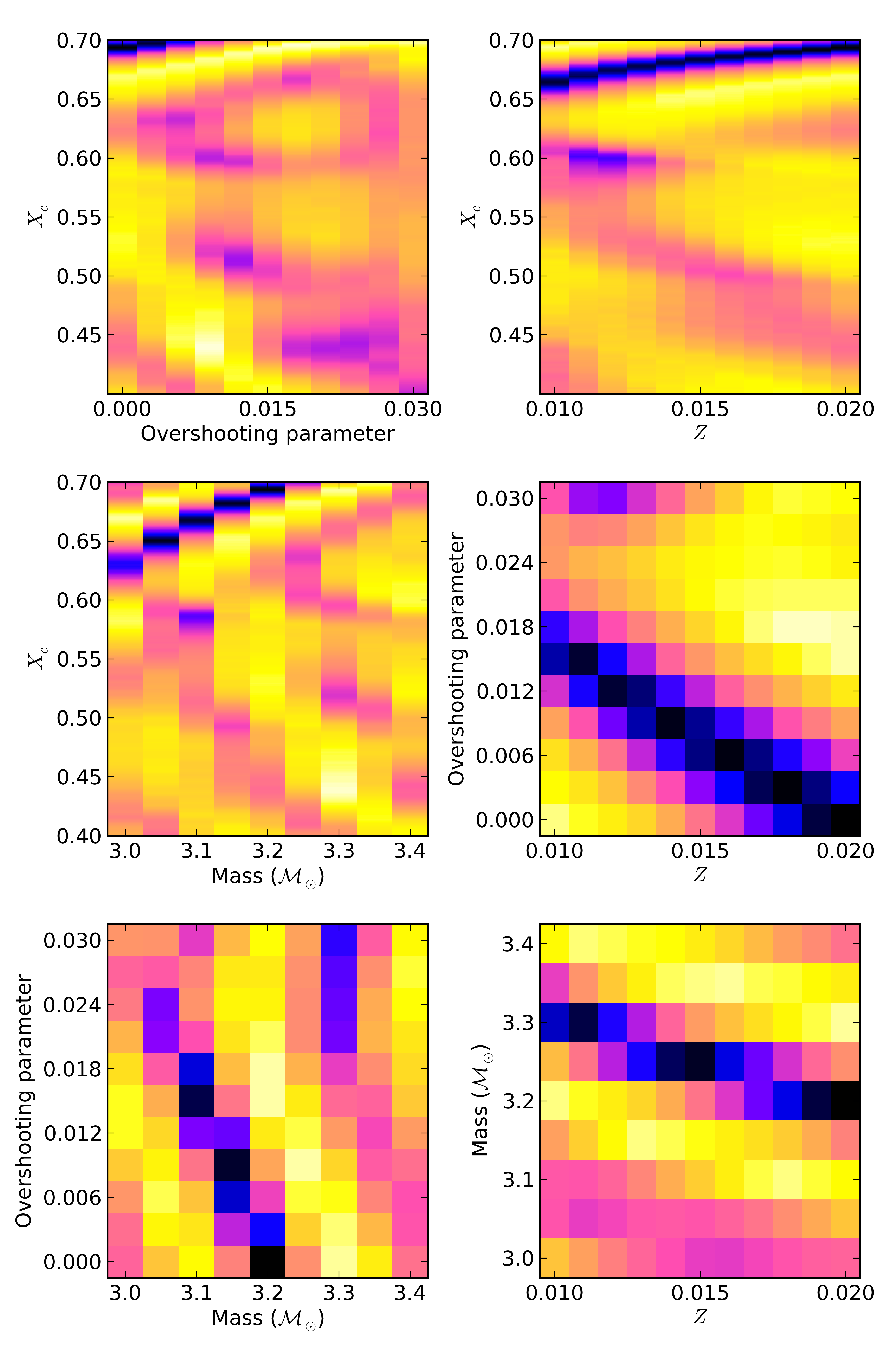}} 
\caption{Visualisation of the correlations between all possible pairs of the four free model parameters within the highly structured $\chi^2$ space. For each subplot, two of the four parameters were fixed to the value corresponding to the best fit model (number 1 in Table\,\ref{bestfitmodles}), and the $\chi^2$ values are plotted in the remaining two-dimensional parameter space using the same colour coding as Fig.\,\ref{MESAchi2}. For further details, see explanation in the text.}
\label{MESAchi2correlations}
\end{figure}

\begin{figure}
\resizebox{\hsize}{!}{\includegraphics{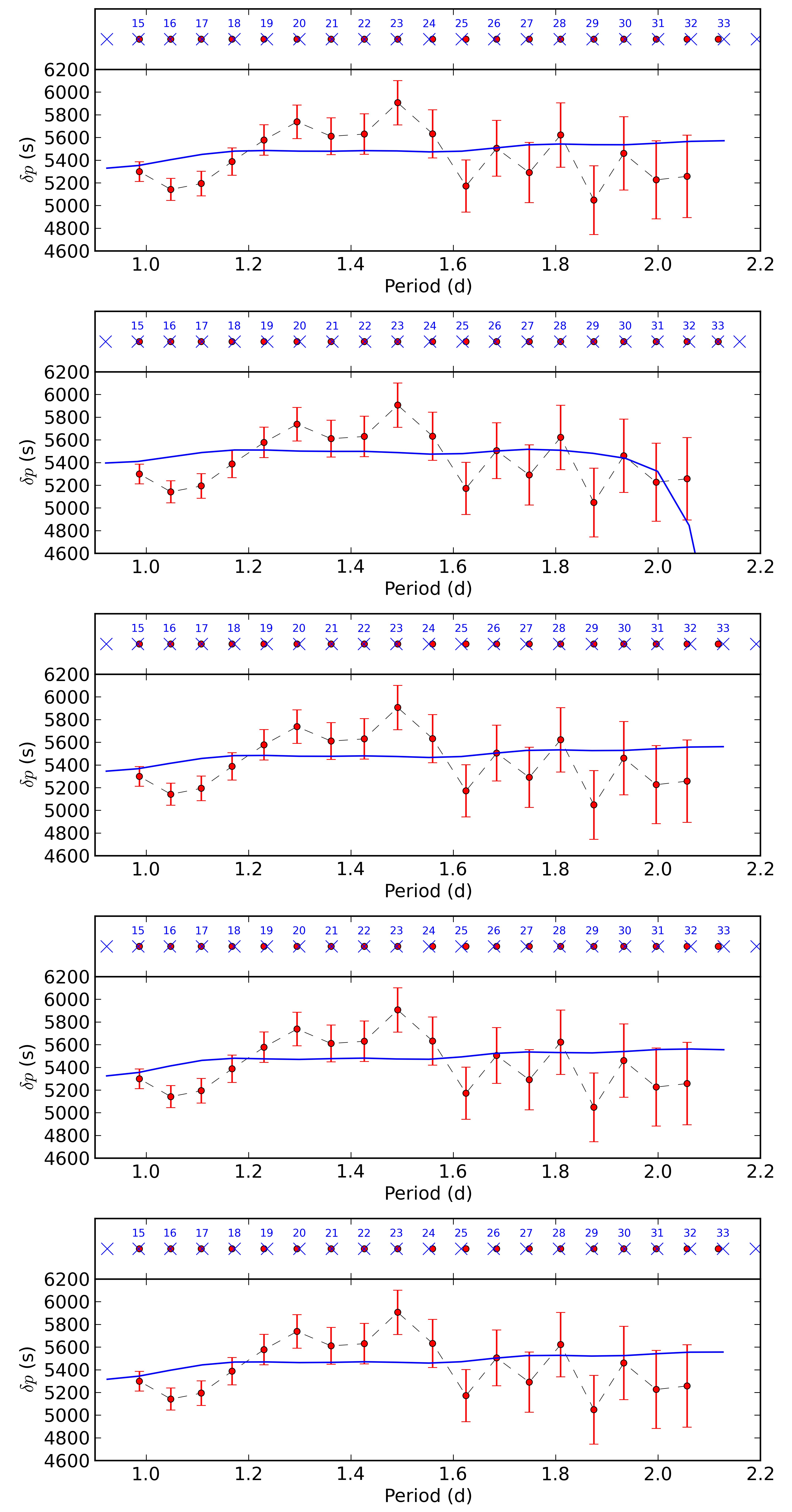}}
\caption{Five best fit models (from top to bottom, following the order in Table\,\ref{bestfitmodles}) using frequencies from MESA/GYRE. (\textit{Upper panels}) Observed peaks (red circles, the error bars are smaller than the symbol size) versus the theoretical frequencies of the $\ell=1$ gravity modes (blue crosses, with the corresponding $n$ value printed above). (\textit{Lower panels}) Individual spacing values between the members of the observed period series (red filled circles) and the $\delta p(p)$ function from the model (blue solid line).}
\label{bestGYREfit}
\end{figure}

\begin{figure}
\resizebox{\hsize}{!}{\includegraphics{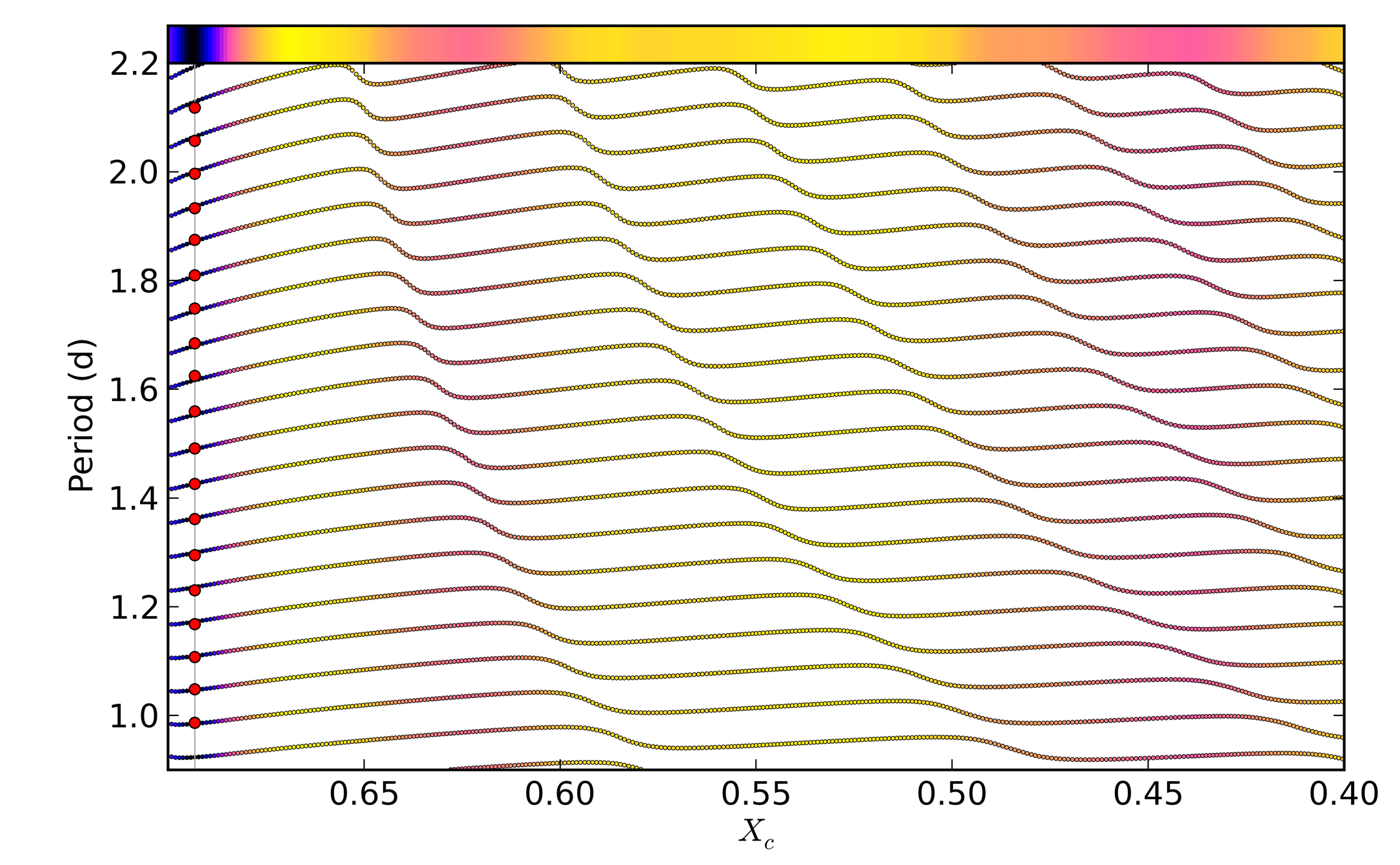}} 
\caption{Evolution of the period of the $\ell=1$ gravity modes with decreasing central hydrogen content. The theoretical frequencies were calculated with GYRE in our region of interest along the MESA evolutionary track which holds the best fitting model ($\mathcal{M}=3.20\mathcal{M}_\odot$, $Z=0.020$, $f_{ov}=0.000$). The colour of each model (a set of GYRE frequencies along a vertical line) represents the logarithm of the corresponding $\chi^2$ value at the given $X_c$ (using the same colours as Fig.\,\ref{MESAchi2}). For better visibility, this colour sequence is repeated on the top of the figure. The observed modes are plotted using larger red circles at the position of the best fitting model at $X_c=0.6932$ (the error bars on the observed values are smaller than the size of the symbols).}
\label{evolutionofGYREfrequencies}
\end{figure}

The $\chi^2$-distributions of stellar parameters from the grid are shown in Fig.\,\ref{MESAchi2}. A first conclusion is that the theoretical frequencies are an adequate representation of the measured ones. It can be noticed that the measured period spacing puts tight constraints on the acceptable models in terms of the radius and age. The effective temperature, mass, overshooting parameter, and metallicity are only moderately constrained, which comes mainly from strong correlations within the model parameters (which is clearly visible on Fig.\,\ref{MESAchi2correlations}) and also from the fact that we cannot constrain metallicity from spectroscopy. The best models are listed in Table\,\ref{bestfitmodles}, while the quality of the best fits to the observed modes and $\delta p(p)$ function is shown in Fig.\,\ref{bestGYREfit}. We show the evolution of the detected $\ell=1$ modes along the evolutionary track, which contains one of the best fit models on Fig.\,\ref{evolutionofGYREfrequencies} for illustrating the high temporal resolution of the grid. 

The correspondence between the observed dipole pulsation frequencies and those predicted by the models is good, if one keeps in mind that we considered a particular choice of input physics for the MESA model computations. Nevertheless, we face limitations in the fine-tuning of the input physics of the models seismically, because the metallicity is too poorly constrained. Indeed, if we were able to fix the metallicity from spectroscopy (take, e.g., $Z=0.016$ without the error bars from Table\,\ref{fundparams}), that would immediately transform into a factor two improvement for all parameters, except core overshooting and $X_c$. This would result in small asteroseismic uncertainties for the parameters other than the core overshooting, for which we can only give an upper limit of $f_{ov}\le0.015$. We note in passing that this corresponds to a step-wise overshoot parameter of $\alpha_{ov}\le0.15$, as hitherto used mostly in seismic modelling \citep{2013EAS....64..323A}. The uncertainty comes from the fact that, in the early stellar evolution, overshooting has little impact on the period spacing. The reason is, that the $\mu$-gradient zones are not developed for such a young star, and the $\delta p(p)$ function exhibits no significant deviation from asymptotic period spacing. From Fig.\,\ref{MESAchi2correlations} it is also clear that we cannot capture all changes smoothly with the current resolution of the grid in mass, since fixing all parameters except mass and moving one step along the mass axis results in quite a large $\chi^2$ change, such that the values in between cannot be interpolated within existing neighbouring model points. Even so, in this specific case of a fairly young star, it is not worth the effort to raise the resolution in mass further, since an enhanced resolution will not provide smaller uncertainties, because overshooting cannot be well constrained (irrespective of the resolution in mass), and the overshooting and mass parameters near the best fit model are strongly correlated. This means that the internal structure in the region where the observed $g$ modes have probing power differ insufficiently in the region where our models provide a good fit, even if this region is quite broad for some of the parameters.

\begin{figure}
\resizebox{\hsize}{!}{\includegraphics{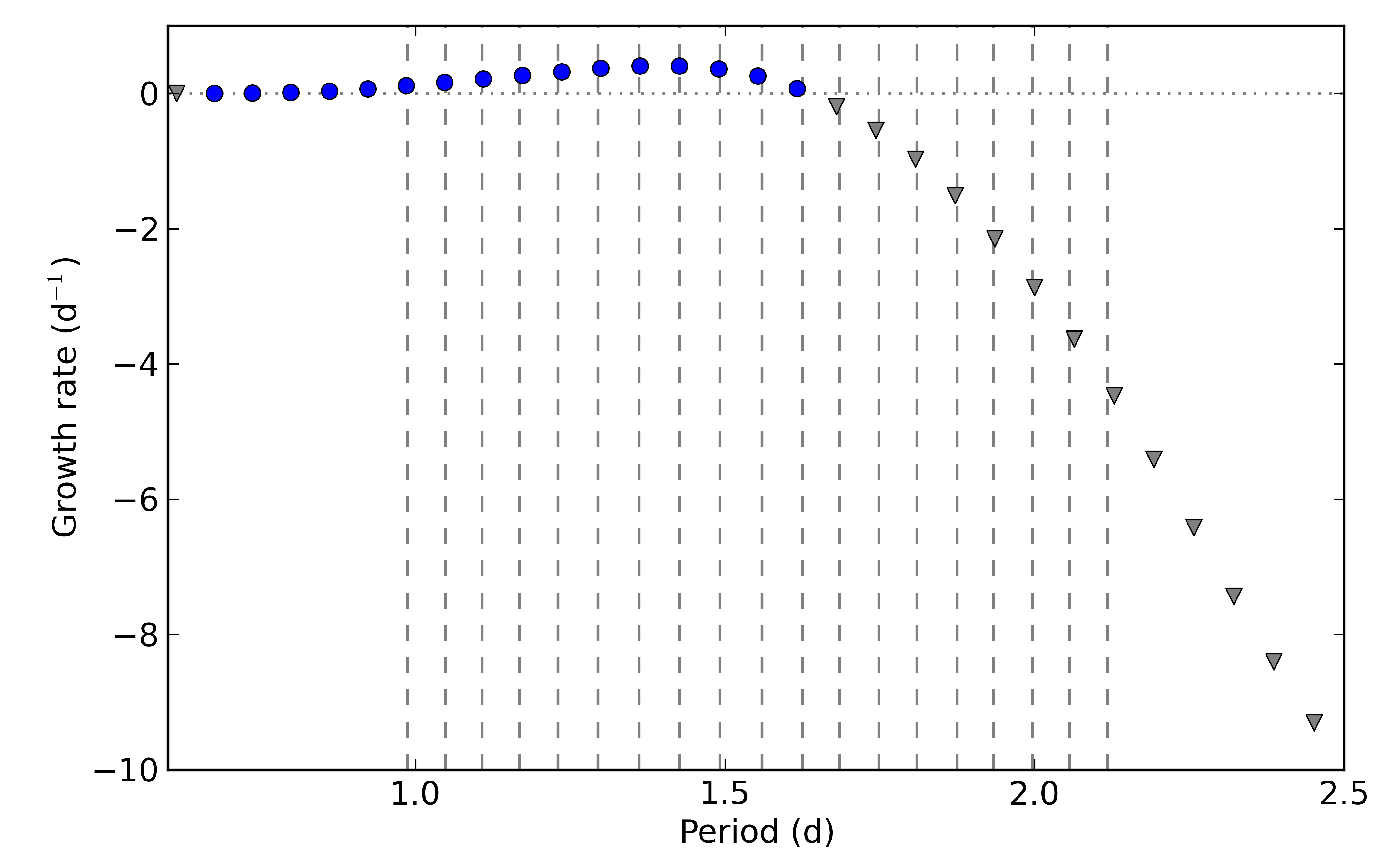}} 
\caption{Non-adiabatic stability analysis for the best fitting model. The growth rate of the zonal dipole modes, as defined in \citet{2010aste.book.....A}, is plotted as a function of the mode period. Excited modes have positive growth rate and are plotted using blue circles, while stable modes are marked with grey triangles. The position of the observed modes are plotted using grey dashed lines.}
\label{GYREnonad}
\end{figure}

Following the modelling discussed above, we subsequently carried out a non-adiabatic analysis to check the dipole mode excitation of the five best models. We explicitly chose not to rely on excitation results prior to the start of the seismic modelling, because it is well known that non-adiabatic excitation computations are not sufficiently reliable to do so, in the sense that some of the detected and identified modes in massive stars are predicted to be stable \citep[e.g.,][]{2011A&A...527A.112B,2011A&A...528A.123P}. Moreover, in contrast to the modelling results, the excitation computations are very dependent on the choice of the opacities \citep{2007MNRAS.375L..21M}. As shown in Fig.\,\ref{GYREnonad} for the best model as an example, more than half of the detected modes are predicted to be excited in the best models. Keeping in mind that we used OPAL opacities for our model grid, which lead to less excited modes than, say, OP opacities, it is reassuring that the selected models have excited modes in the appropriate frequency range.

As far as we are aware, this is the first seismic modelling of a massive star done with MESA/GYRE. The results are excellent, given that we could rely only on a series of high-order $g$ modes in a very young star, restricted to dipole modes.

\begin{table*}
\caption{Parameters of the five best fitting models. See text for details.}
\label{bestfitmodles}
\centering
\renewcommand{\arraystretch}{1.1}
\begin{tabular}{c c c c c c c c c c c c}
\hline\hline
Model&$T_\mathrm{eff}$&$\log g$&Mass&Radius&Core overshoot ($f_{ov}$)&$Z$&$X_c$&Age&$\chi^2_{spacings}$&$\chi^2_{periods}$&$\chi^{\mathrm{2}}$\\
&K&dex&$\mathcal{M}_\odot$&$R_\odot$&&\multicolumn{2}{c}{mass fraction}&Myr&&&\\
\hline
1&12\,470&4.30&3.20&2.10&0.000&0.020&0.693&12.0&2.213&10.921&6.567\\
2&11\,760&4.27&3.00&2.11&0.015&0.020&0.665&45.1&2.500&10.797&6.649\\
3&12\,310&4.30&3.15&2.09&0.006&0.020&0.690&15.8&2.305&11.183&6.744\\
4&13\,140&4.33&3.25&2.04&0.000&0.016&0.696&10.5&2.337&11.178&6.757\\
5&12\,610&4.31&3.20&2.07&0.003&0.019&0.695&11.2&2.204&11.382&6.793\\
\hline
\end{tabular}
\end{table*}

\section{Conclusions}
We have presented the first detailed asteroseismic analysis of a single, main sequence B-type star based on four years of \textit{Kepler} photometry. It led to the classification of KIC\,10526294 as a very slowly rotating SPB star, and -- most importantly -- the discovery of a series of nineteen quasi-equally spaced dipole modes, each showing a very narrow rotationally split triplet structure. The observed splittings are systematically higher towards longer periods at a level that points towards a non-rigid internal rotation profile. The rotational splitting for the retrograde and prograde modes is slightly asymmetric.

The result of the seismic modelling of the ($\ell=1$, $m=0$) dipole series for a dense model grid with evolutionary tracks covering the spectroscopic error box indicates that the star is young ($X_c>0.64$). Owing to the numerous correlations between various stellar parameters at such early evolutionary stage and to the high uncertainty on the observed metallicity, we can only place an upper limit for the extent of the core overshooting region ($\alpha_{ov}\le0.15$), yet this is a stringent seismic constraint.

Our work represents the third detection of a series of quasi-equally spaced gravity modes in a main sequence B type star \citep[after][]{2010Natur.464..259D,2012A&A...542A..55P}, but this is the first time that such a long series has been detected. Moreover, all the rotationally split components were observed and leave no doubt that we are dealing with dipole modes. This mode identification, which was an {\it assumption\/} in the model comparison done for HD\,50230 by \citet{2010Natur.464..259D}, implies that we could achieve the first actual seismic modelling of an SPB star.

In a similar type of analysis, \citet{2014arXiv1405.0155K} found a series of equally spaced and rotationally split gravity-mode triplets in the slowly rotating A-type pulsator KIC\,11145123, also from four years of \textit{Kepler} photometry. In addition, this star exhibits rotationally split $p$ modes. While the authors did not perform spectroscopic observations or seismic modelling from the zonal modes detected in this star, the \textit{Kepler} data itself was rich enough in terms of frequency content to conclude that this star 
has a rotation period of some 100\,d and a faster envelope than core rotation.

In the next step, we plan to investigate if additional acceptable models occur in a much more extended model grid that does not start from the observed spectroscopic $T_\mathrm{eff}$ and $\log g$. Moreover, we shall add one dimension into the grid by exploring whether the measured oscillatory properties allow the distinction between core overshooting as we treated it here and the effect of semi-convective mixing. Finally, we plan to use the kernels for the dipole modes of the best models to perform a frequency inversion of the rotationally split frequencies with the aim of deriving the internal rotation profile of the star. Our first attempts already allow the conclusion 
that the rotation profile cannot be constant throughout the star.


\begin{acknowledgements}
The research leading to these results has received funding from the Research Fund KU Leuven, from the European Research Council under the European Community's Seventh Framework Programme (FP7/2007--2013)/ERC grant agreement n$^\circ$227224 (PROSPERITY), as well as from the Belgian Science Policy Office (Belspo, C90309: CoRoT Data Exploitation). For the model grid computations, we made use of the infrastructure of the VSC -- Flemish Supercomputer Center, funded by the Hercules Foundation and the Flemish Government -- department EWI. EM and CA are grateful to Bill Paxton, Rich Townsend, and the entire MESA/GYRE development team for their efforts and free distribution of their software, as well as to the staff of the Kavli Institute of Theoretical Physics, UCSB, for the kind hospitality during the MESA 2013 Summer School, which was a memorable experience. SB is supported by the Foundation for Fundamental Research on Matter (FOM), which is part of the Netherlands Organisation for Scientific Research (NWO). Funding for the \textit{Kepler} mission is provided by the NASA Science Mission directorate. Some of the data presented in this paper were obtained from the Multimission Archive at the Space Telescope Science Institute (MAST). STScI is operated by the Association of Universities for Research in Astronomy, Inc., under NASA contract NAS5-26555. Support for MAST for non-HST data is provided by the NASA Office of Space Science via grant NNX09AF08G and by other grants and contracts. This work is partly based on observations made with the William Herschel Telescope, which is operated on the island of La Palma by the Isaac Newton Group in the Spanish Observatorio del Roque de los Muchachos (ORM) of the Instituto de Astrof\'{i}sica de Canarias (IAC).
\end{acknowledgements}

\bibliographystyle{aa}
\bibliography{KeplerBstars_II}

\begin{thebibliography}{53}
\expandafter\ifx\csname natexlab\endcsname\relax\def\natexlab#1{#1}\fi

\bibitem[{{Aerts}(2013)}]{2013EAS....64..323A}
{Aerts}, C. 2013, in EAS Publications Series, Vol.~64, EAS Publications Series,
  323--330

\bibitem[{{Aerts} {et~al.}(2010){Aerts}, {Christensen-Dalsgaard}, \&
  {Kurtz}}]{2010aste.book.....A}
{Aerts}, C., {Christensen-Dalsgaard}, J., \& {Kurtz}, D.~W. 2010,
  {Asteroseismology} ({Astronomy and Astrophsyics Library, Springer Berlin
  Heidelberg})

\bibitem[{{Aerts} {et~al.}(2003){Aerts}, {Thoul}, {Daszy{\'n}ska}, {Scuflaire},
  {Waelkens}, {Dupret}, {Niemczura}, \& {Noels}}]{2003Sci...300.1926A}
{Aerts}, C., {Thoul}, A., {Daszy{\'n}ska}, J., {et~al.} 2003, Science, 300,
  1926

\bibitem[{{Aerts} {et~al.}(2004){Aerts}, {Waelkens},
  {Daszy{\'n}ska-Daszkiewicz}, {Dupret}, {Thoul}, {Scuflaire}, {Uytterhoeven},
  {Niemczura}, \& {Noels}}]{2004A&A...415..241A}
{Aerts}, C., {Waelkens}, C., {Daszy{\'n}ska-Daszkiewicz}, J., {et~al.} 2004,
  \aap, 415, 241

\bibitem[{{Asplund} {et~al.}(2005){Asplund}, {Grevesse}, \&
  {Sauval}}]{2005ASPC..336...25A}
{Asplund}, M., {Grevesse}, N., \& {Sauval}, A.~J. 2005, in Astronomical Society
  of the Pacific Conference Series, Vol. 336, Cosmic Abundances as Records of
  Stellar Evolution and Nucleosynthesis, ed. {T.~G.~Barnes III \& F.~N.~Bash},
  25

\bibitem[{{Auvergne} {et~al.}(2009){Auvergne}, {Bodin}, {Boisnard}, {Buey},
  {Chaintreuil}, {Epstein}, {Jouret}, {Lam-Trong}, {Levacher}, {Magnan},
  {Perez}, {Plasson}, {Plesseria}, {Peter}, {Steller}, {Tiph{\`e}ne}, {Baglin},
  {Agogu{\'e}}, {Appourchaux}, {Barbet}, {Beaufort}, {Bellenger}, {Berlin},
  {Bernardi}, {Blouin}, {Boumier}, {Bonneau}, {Briet}, {Butler}, {Cautain},
  {Chiavassa}, {Costes}, {Cuvilho}, {Cunha-Parro}, {de Oliveira Fialho},
  {Decaudin}, {Defise}, {Djalal}, {Docclo}, {Drummond}, {Dupuis}, {Exil},
  {Faur{\'e}}, {Gaboriaud}, {Gamet}, {Gavalda}, {Grolleau}, {Gueguen},
  {Guivarc'h}, {Guterman}, {Hasiba}, {Huntzinger}, {Hustaix}, {Imbert},
  {Jeanville}, {Johlander}, {Jorda}, {Journoud}, {Karioty}, {Kerjean},
  {Lafond}, {Lapeyrere}, {Landiech}, {Larqu{\'e}}, {Laudet}, {Le Merrer},
  {Leporati}, {Leruyet}, {Levieuge}, {Llebaria}, {Martin}, {Mazy}, {Mesnager},
  {Michel}, {Moalic}, {Monjoin}, {Naudet}, {Neukirchner}, {Nguyen-Kim},
  {Ollivier}, {Orcesi}, {Ottacher}, {Oulali}, {Parisot}, {Perruchot},
  {Piacentino}, {Pinheiro da Silva}, {Platzer}, {Pontet}, {Pradines},
  {Quentin}, {Rohbeck}, {Rolland}, {Rollenhagen}, {Romagnan}, {Russ}, {Samadi},
  {Schmidt}, {Schwartz}, {Sebbag}, {Smit}, {Sunter}, {Tello}, {Toulouse},
  {Ulmer}, {Vandermarcq}, {Vergnault}, {Wallner}, {Waultier}, \&
  {Zanatta}}]{2009A&A...506..411A}
{Auvergne}, M., {Bodin}, P., {Boisnard}, L., {et~al.} 2009, \aap, 506, 411

\bibitem[{{Balona} {et~al.}(2011){Balona}, {Pigulski}, {De Cat}, {Handler},
  {Guti{\'e}rrez-Soto}, {Engelbrecht}, {Frescura}, {Briquet}, {Cuypers},
  {Daszy{\'n}ska-Daszkiewicz}, {Degroote}, {Dukes}, {Garcia}, {Green}, {Heber},
  {Kawaler}, {Lehmann}, {Leroy}, {Molenda-{\.Z}aaowicz}, {Neiner}, {Noels},
  {Nuspl}, {{\O}stensen}, {Pricopi}, {Roxburgh}, {Salmon}, {Smith},
  {Su{\'a}rez}, {Suran}, {Szab{\'o}}, {Uytterhoeven}, {Christensen-Dalsgaard},
  {Kjeldsen}, {Caldwell}, {Girouard}, \& {Sanderfer}}]{2011MNRAS.413.2403B}
{Balona}, L.~A., {Pigulski}, A., {De Cat}, P., {et~al.} 2011, \mnras, 413, 2403

\bibitem[{{Borucki} {et~al.}(2010){Borucki}, {Koch}, {Basri}, {Batalha},
  {Brown}, {Caldwell}, {Caldwell}, {Christensen-Dalsgaard}, {Cochran},
  {DeVore}, {Dunham}, {Dupree}, {Gautier}, {Geary}, {Gilliland}, {Gould},
  {Howell}, {Jenkins}, {Kondo}, {Latham}, {Marcy}, {Meibom}, {Kjeldsen},
  {Lissauer}, {Monet}, {Morrison}, {Sasselov}, {Tarter}, {Boss}, {Brownlee},
  {Owen}, {Buzasi}, {Charbonneau}, {Doyle}, {Fortney}, {Ford}, {Holman},
  {Seager}, {Steffen}, {Welsh}, {Rowe}, {Anderson}, {Buchhave}, {Ciardi},
  {Walkowicz}, {Sherry}, {Horch}, {Isaacson}, {Everett}, {Fischer}, {Torres},
  {Johnson}, {Endl}, {MacQueen}, {Bryson}, {Dotson}, {Haas}, {Kolodziejczak},
  {Van Cleve}, {Chandrasekaran}, {Twicken}, {Quintana}, {Clarke}, {Allen},
  {Li}, {Wu}, {Tenenbaum}, {Verner}, {Bruhweiler}, {Barnes}, \&
  {Prsa}}]{2010Sci...327..977B}
{Borucki}, W.~J., {Koch}, D., {Basri}, G., {et~al.} 2010, Science, 327, 977

\bibitem[{{Breger} {et~al.}(1993){Breger}, {Stich}, {Garrido}, {Martin},
  {Jiang}, {Li}, {Hube}, {Ostermann}, {Paparo}, \&
  {Scheck}}]{1993A&A...271..482B}
{Breger}, M., {Stich}, J., {Garrido}, R., {et~al.} 1993, \aap, 271, 482

\bibitem[{{Briquet} {et~al.}(2011){Briquet}, {Aerts}, {Baglin}, {Nieva},
  {Degroote}, {Przybilla}, {Noels}, {Schiller}, {Vu{\v c}kovi{\'c}}, {Oreiro},
  {Smolders}, {Auvergne}, {Baudin}, {Catala}, {Michel}, \&
  {Samadi}}]{2011A&A...527A.112B}
{Briquet}, M., {Aerts}, C., {Baglin}, A., {et~al.} 2011, \aap, 527, A112

\bibitem[{{Chaplin} {et~al.}(2013){Chaplin}, {Sanchis-Ojeda}, {Campante},
  {Handberg}, {Stello}, {Winn}, {Basu}, {Christensen-Dalsgaard}, {Davies},
  {Metcalfe}, {Buchhave}, {Fischer}, {Bedding}, {Cochran}, {Elsworth},
  {Gilliland}, {Hekker}, {Huber}, {Isaacson}, {Karoff}, {Kawaler}, {Kjeldsen},
  {Latham}, {Lund}, {Lundkvist}, {Marcy}, {Miglio}, {Barclay}, \&
  {Lissauer}}]{2013ApJ...766..101C}
{Chaplin}, W.~J., {Sanchis-Ojeda}, R., {Campante}, T.~L., {et~al.} 2013, \apj,
  766, 101

\bibitem[{{Cutri} {et~al.}(2003){Cutri}, {Skrutskie}, {van Dyk}, {Beichman},
  {Carpenter}, {Chester}, {Cambresy}, {Evans}, {Fowler}, {Gizis}, {Howard},
  {Huchra}, {Jarrett}, {Kopan}, {Kirkpatrick}, {Light}, {Marsh}, {McCallon},
  {Schneider}, {Stiening}, {Sykes}, {Wheaton}, {Wheelock}, \&
  {Zacarias}}]{2mass}
{Cutri}, R.~M., {Skrutskie}, M.~F., {van Dyk}, S., {et~al.} 2003, {2MASS All
  Sky Catalog of point sources.}

\bibitem[{{De Cat} {et~al.}(2005){De Cat}, {Briquet},
  {Daszy{\'n}ska-Daszkiewicz}, {Dupret}, {De Ridder}, {Scuflaire}, \&
  {Aerts}}]{2005A&A...432.1013D}
{De Cat}, P., {Briquet}, M., {Daszy{\'n}ska-Daszkiewicz}, J., {et~al.} 2005,
  \aap, 432, 1013

\bibitem[{{Debosscher} {et~al.}(2011){Debosscher}, {Blomme}, {Aerts}, \& {De
  Ridder}}]{2011A&A...529A..89D}
{Debosscher}, J., {Blomme}, J., {Aerts}, C., \& {De Ridder}, J. 2011, \aap,
  529, A89

\bibitem[{{Degroote}(2010)}]{2010_degroote_phd}
{Degroote}, P. 2010, PhD thesis, KU Leuven, Belgium

\bibitem[{{Degroote} {et~al.}(2010){Degroote}, {Aerts}, {Baglin}, {Miglio},
  {Briquet}, {Noels}, {Niemczura}, {Montalban}, {Bloemen}, {Oreiro}, {Vu{\v
  c}kovi{\'c}}, {Smolders}, {Auvergne}, {Baudin}, {Catala}, \&
  {Michel}}]{2010Natur.464..259D}
{Degroote}, P., {Aerts}, C., {Baglin}, A., {et~al.} 2010, \nat, 464, 259

\bibitem[{{Degroote} {et~al.}(2012){Degroote}, {Aerts}, {Michel}, {Briquet},
  {P{\'a}pics}, {Amado}, {Mathias}, {Poretti}, {Rainer}, {Lombaert}, {Hillen},
  {Morel}, {Auvergne}, {Baglin}, {Baudin}, {Catala}, \&
  {Samadi}}]{2012A&A..542A..88D}
{Degroote}, P., {Aerts}, C., {Michel}, E., {et~al.} 2012, \aap, 542, A88

\bibitem[{{Degroote} {et~al.}(2009){Degroote}, {Briquet}, {Catala},
  {Uytterhoeven}, {Lefever}, {Morel}, {Aerts}, {Carrier}, {Auvergne}, {Baglin},
  \& {Michel}}]{2009A&A...506..111D}
{Degroote}, P., {Briquet}, M., {Catala}, C., {et~al.} 2009, \aap, 506, 111

\bibitem[{{Dupret} {et~al.}(2004){Dupret}, {Thoul}, {Scuflaire},
  {Daszy{\'n}ska-Daszkiewicz}, {Aerts}, {Bourge}, {Waelkens}, \&
  {Noels}}]{2004A&A...415..251D}
{Dupret}, M.-A., {Thoul}, A., {Scuflaire}, R., {et~al.} 2004, \aap, 415, 251

\bibitem[{{Dziembowski}(1977)}]{1977AcA....27...95D}
{Dziembowski}, W. 1977, \actaa, 27, 95

\bibitem[{{Gabriel} {et~al.}(2014){Gabriel}, {Noels}, {Montalban}, \&
  {Miglio}}]{2014arXiv1405.0128G}
{Gabriel}, M., {Noels}, A., {Montalban}, J., \& {Miglio}, A. 2014, ArXiv
  1405.0128

\bibitem[{{Gilliland} {et~al.}(2010){Gilliland}, {Brown},
  {Christensen-Dalsgaard}, {Kjeldsen}, {Aerts}, {Appourchaux}, {Basu},
  {Bedding}, {Chaplin}, {Cunha}, {De Cat}, {De Ridder}, {Guzik}, {Handler},
  {Kawaler}, {Kiss}, {Kolenberg}, {Kurtz}, {Metcalfe}, {Monteiro}, {Szab{\'o}},
  {Arentoft}, {Balona}, {Debosscher}, {Elsworth}, {Quirion}, {Stello},
  {Su{\'a}rez}, {Borucki}, {Jenkins}, {Koch}, {Kondo}, {Latham}, {Rowe}, \&
  {Steffen}}]{2010PASP..122..131G}
{Gilliland}, R.~L., {Brown}, T.~M., {Christensen-Dalsgaard}, J., {et~al.} 2010,
  \pasp, 122, 131

\bibitem[{{Herwig}(2000)}]{2000A&A...360..952H}
{Herwig}, F. 2000, \aap, 360, 952

\bibitem[{{Iglesias} \& {Rogers}(1996)}]{1996ApJ...464..943I}
{Iglesias}, C.~A. \& {Rogers}, F.~J. 1996, \apj, 464, 943

\bibitem[{{\textit{Kepler} Mission Team}(2009)}]{2009yCat.5133....0K}
{\textit{Kepler} Mission Team}. 2009, VizieR Online Data Catalog, 5133, 0

\bibitem[{{Kippenhahn} {et~al.}(2013){Kippenhahn}, {Weigert}, \&
  {Weiss}}]{2013sse..book.....K}
{Kippenhahn}, R., {Weigert}, A., \& {Weiss}, A. 2013, {Stellar Structure and
  Evolution}

\bibitem[{{Kurtz} {et~al.}(2014){Kurtz}, {Saio}, {Takata}, {Shibahashi},
  {Murphy}, \& {Sekii}}]{2014arXiv1405.0155K}
{Kurtz}, D.~W., {Saio}, H., {Takata}, M., {et~al.} 2014, \mnras, in press,
  ArXiv 1405.0155

\bibitem[{{Langer} {et~al.}(1983){Langer}, {Fricke}, \&
  {Sugimoto}}]{1983A&A...126..207L}
{Langer}, N., {Fricke}, K.~J., \& {Sugimoto}, D. 1983, \aap, 126, 207

\bibitem[{{Lehmann} {et~al.}(2011){Lehmann}, {Tkachenko}, {Semaan},
  {Guti{\'e}rrez-Soto}, {Smalley}, {Briquet}, {Shulyak}, {Tsymbal}, \& {De
  Cat}}]{2011A&A...526A.124L}
{Lehmann}, H., {Tkachenko}, A., {Semaan}, T., {et~al.} 2011, \aap, 526, A124

\bibitem[{{Loumos} \& {Deeming}(1978)}]{1978Ap&SS..56..285L}
{Loumos}, G.~L. \& {Deeming}, T.~J. 1978, \apss, 56, 285

\bibitem[{{Marsh}(1989)}]{1989PASP..101.1032M}
{Marsh}, T.~R. 1989, \pasp, 101, 1032

\bibitem[{{McNamara} {et~al.}(2012){McNamara}, {Jackiewicz}, \&
  {McKeever}}]{2012AJ....143..101M}
{McNamara}, B.~J., {Jackiewicz}, J., \& {McKeever}, J. 2012, \aj, 143, 101

\bibitem[{{Miglio} {et~al.}(2007){Miglio}, {Montalb{\'a}n}, \&
  {Dupret}}]{2007MNRAS.375L..21M}
{Miglio}, A., {Montalb{\'a}n}, J., \& {Dupret}, M.-A. 2007, \mnras, 375, L21

\bibitem[{{Miglio} {et~al.}(2008){Miglio}, {Montalb{\'a}n}, {Noels}, \&
  {Eggenberger}}]{2008MNRAS.386.1487M}
{Miglio}, A., {Montalb{\'a}n}, J., {Noels}, A., \& {Eggenberger}, P. 2008,
  \mnras, 386, 1487

\bibitem[{{Monet} {et~al.}(2003){Monet}, {Levine}, {Canzian}, {Ables}, {Bird},
  {Dahn}, {Guetter}, {Harris}, {Henden}, {Leggett}, {Levison}, {Luginbuhl},
  {Martini}, {Monet}, {Munn}, {Pier}, {Rhodes}, {Riepe}, {Sell}, {Stone},
  {Vrba}, {Walker}, {Westerhout}, {Brucato}, {Reid}, {Schoening}, {Hartley},
  {Read}, \& {Tritton}}]{2003AJ....125..984M}
{Monet}, D.~G., {Levine}, S.~E., {Canzian}, B., {et~al.} 2003, \aj, 125, 984

\bibitem[{{Nieva} \& {Przybilla}(2012)}]{2012A&A...539A.143N}
{Nieva}, M.-F. \& {Przybilla}, N. 2012, \aap, 539, A143

\bibitem[{{Pamyatnykh}(1999)}]{1999AcA....49..119P}
{Pamyatnykh}, A.~A. 1999, \actaa, 49, 119

\bibitem[{{P{\'a}pics}(2012)}]{2012AN....333.1053P}
{P{\'a}pics}, P.~I. 2012, Astronomische Nachrichten, 333, 1053

\bibitem[{{P{\'a}pics}(2013)}]{papics_phd}
{P{\'a}pics}, P.~I. 2013, PhD thesis, KU Leuven, Belgium

\bibitem[{{P{\'a}pics} {et~al.}(2011){P{\'a}pics}, {Briquet}, {Auvergne},
  {Aerts}, {Degroote}, {Niemczura}, {Vu{\v c}kovi{\'c}}, {Smolders}, {Poretti},
  {Rainer}, {Hareter}, {Baglin}, {Baudin}, {Catala}, {Michel}, \&
  {Samadi}}]{2011A&A...528A.123P}
{P{\'a}pics}, P.~I., {Briquet}, M., {Auvergne}, M., {et~al.} 2011, \aap, 528,
  A123

\bibitem[{{P{\'a}pics} {et~al.}(2012){P{\'a}pics}, {Briquet}, {Baglin},
  {Poretti}, {Aerts}, {Degroote}, {Tkachenko}, {Morel}, {Zima}, {Niemczura},
  {Rainer}, {Hareter}, {Baudin}, {Catala}, {Michel}, {Samadi}, \&
  {Auvergne}}]{2012A&A...542A..55P}
{P{\'a}pics}, P.~I., {Briquet}, M., {Baglin}, A., {et~al.} 2012, \aap, 542, A55

\bibitem[{{P{\'a}pics} {et~al.}(2013){P{\'a}pics}, {Tkachenko}, {Aerts},
  {Briquet}, {Marcos-Arenal}, {Beck}, {Uytterhoeven}, {Trivi{\~n}o Hage},
  {Southworth}, {Clubb}, {Bloemen}, {Degroote}, {Jackiewicz}, {McKeever}, {Van
  Winckel}, {Niemczura}, {Gameiro}, \& {Debosscher}}]{2013A&A...553A.127P}
{P{\'a}pics}, P.~I., {Tkachenko}, A., {Aerts}, C., {et~al.} 2013, \aap, 553,
  A127

\bibitem[{{Paxton} {et~al.}(2011){Paxton}, {Bildsten}, {Dotter}, {Herwig},
  {Lesaffre}, \& {Timmes}}]{2011ApJS..192....3P}
{Paxton}, B., {Bildsten}, L., {Dotter}, A., {et~al.} 2011, \apjs, 192, 3

\bibitem[{{Paxton} {et~al.}(2013){Paxton}, {Cantiello}, {Arras}, {Bildsten},
  {Brown}, {Dotter}, {Mankovich}, {Montgomery}, {Stello}, {Timmes}, \&
  {Townsend}}]{2013ApJS..208....4P}
{Paxton}, B., {Cantiello}, M., {Arras}, P., {et~al.} 2013, \apjs, 208, 4

\bibitem[{{Przybilla} {et~al.}(2013){Przybilla}, {Nieva}, {Irrgang}, \&
  {Butler}}]{2013EAS....63...13P}
{Przybilla}, N., {Nieva}, M.~F., {Irrgang}, A., \& {Butler}, K. 2013, in EAS
  Publications Series, Vol.~63, EAS Publications Series, 13--23

\bibitem[{{Scargle}(1982)}]{1982ApJ...263..835S}
{Scargle}, J.~D. 1982, \apj, 263, 835

\bibitem[{{Schmidt-Kaler}(1982)}]{1982SchmidtKalerBook}
{Schmidt-Kaler}, T. 1982, {Landolt-B{\"o}rnstein}, ed. {K.~Schaifers \&
  H.~H.~Vogt}, Vol.~2b ({Springer--Verlag})

\bibitem[{{Seaton}(2005)}]{2005MNRAS.362L...1S}
{Seaton}, M.~J. 2005, \mnras, 362, L1

\bibitem[{{Shulyak} {et~al.}(2004){Shulyak}, {Tsymbal}, {Ryabchikova},
  {St{\"u}tz}, \& {Weiss}}]{2004A&A...428..993S}
{Shulyak}, D., {Tsymbal}, V., {Ryabchikova}, T., {St{\"u}tz}, C., \& {Weiss},
  W.~W. 2004, \aap, 428, 993

\bibitem[{{Tkachenko} {et~al.}(2012){Tkachenko}, {Lehmann}, {Smalley},
  {Debosscher}, \& {Aerts}}]{2012MNRAS.422.2960T}
{Tkachenko}, A., {Lehmann}, H., {Smalley}, B., {Debosscher}, J., \& {Aerts}, C.
  2012, \mnras, 422, 2960

\bibitem[{{Townsend} \& {Teitler}(2013)}]{2013MNRAS.435.3406T}
{Townsend}, R.~H.~D. \& {Teitler}, S.~A. 2013, \mnras, 435, 3406

\bibitem[{{Tsymbal}(1996)}]{1996ASPC..108..198T}
{Tsymbal}, V. 1996, in Astronomical Society of the Pacific Conference Series,
  Vol. 108, M.A.S.S., Model Atmospheres and Spectrum Synthesis, ed.
  {S.~J.~Adelman, F.~Kupka, \& W.~W.~Weiss}, 198

\bibitem[{{Walker} {et~al.}(2003){Walker}, {Matthews}, {Kuschnig}, {Johnson},
  {Rucinski}, {Pazder}, {Burley}, {Walker}, {Skaret}, {Zee}, {Grocott},
  {Carroll}, {Sinclair}, {Sturgeon}, \& {Harron}}]{2003PASP..115.1023W}
{Walker}, G., {Matthews}, J., {Kuschnig}, R., {et~al.} 2003, \pasp, 115, 1023

\end{thebibliography}


\Online
\begin{appendix}

\section{Figures}\label{figures}
\begin{figure*}
\resizebox{\hsize}{!}{\includegraphics{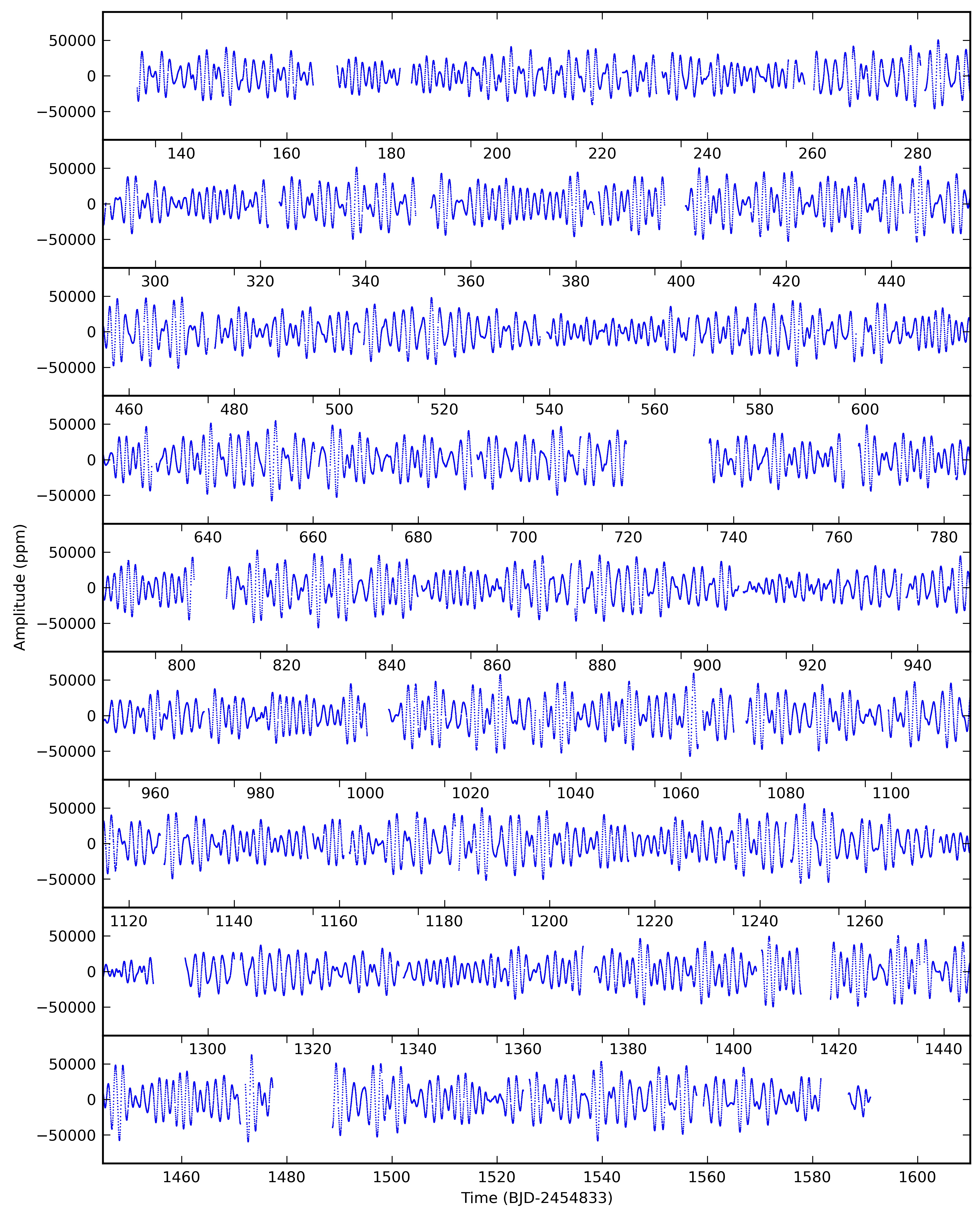}} 
\caption{Full reduced \textit{Kepler} light curve (blue dots, brightest at the top) of KIC\,10526294.}
\label{lightcurve}
\end{figure*}

\section{Tables}\label{tables}
\begin{longtable}{c c c c c c c c c}
\caption{\label{frequtable} Fourier parameters (frequencies ($f_j$), amplitudes ($A_j$), and phases ($\theta_j$)) of the significant peaks having a S/N above 4 for KIC\,10526294 when computed in a $1\,\mathrm{d}^{-1}$ window after prewhitening. The displayed S/N values are calculated in a window of $1\,\mathrm{d}^{-1}$ or $3\,\mathrm{d}^{-1}$ centred on the given frequency, or from the full periodogram (from $0\,\mathrm{d}^{-1}$ to $24.48\,\mathrm{d}^{-1}$). A complete list of all model frequencies is available upon request.}\\
\hline\hline
$f$ & $\epsilon_f$ & $A$ & $\epsilon_A$ & $\theta$ & $\epsilon_{\theta}$ & \multicolumn{3}{c}{S/N}\\
$\mathrm{d}^{-1}$ & $\mathrm{d}^{-1}$ & $\mathrm{ppm}$ & $\mathrm{ppm}$ & $2\pi/\mathrm{rad}$ & $2\pi/\mathrm{rad}$ & $1\,\mathrm{d}^{-1}$ & $3\,\mathrm{d}^{-1}$ & full\\
\hline
\endfirsthead
\caption{continued.}\\
\hline\hline
$f$ & $\epsilon_f$ & $A$ & $\epsilon_A$ & $\theta$ & $\epsilon_{\theta}$ & \multicolumn{3}{c}{S/N}\\
$\mathrm{d}^{-1}$ & $\mathrm{d}^{-1}$ & $\mathrm{ppm}$ & $\mathrm{ppm}$ & $2\pi/\mathrm{rad}$ & $2\pi/\mathrm{rad}$ & $1\,\mathrm{d}^{-1}$ & $3\,\mathrm{d}^{-1}$ & full\\
\hline
\endhead
\hline
\endfoot
  0.002100 &   0.000020 &   131.3 &     6.8 &   0.0313 &   0.0537 &   4.1 &  12.9 &  37.1 \\
  0.006627 &   0.000017 &   193.2 &     8.3 &   0.2402 &   0.0445 &   5.1 &  16.4 &  48.2 \\
  0.008148 &   0.000017 &   189.3 &     8.4 &   0.1798 &   0.0462 &   4.8 &  15.6 &  45.8 \\
  0.009521 &   0.000017 &   198.0 &     8.3 &   0.3139 &   0.0459 &   4.9 &  15.8 &  46.6 \\
  0.018700 &   0.000020 &   136.6 &     6.9 &  -0.4251 &   0.0540 &   4.0 &  12.8 &  36.8 \\
  0.073198 &   0.000022 &    99.9 &     6.0 &   0.4932 &   0.0590 &   5.1 &  11.4 &  32.4 \\
  0.163434 &   0.000021 &   116.8 &     6.8 &   0.1341 &   0.0546 &   4.0 &  12.6 &  36.4 \\
  0.165589 &   0.000016 &   196.6 &     8.8 &  -0.4784 &   0.0435 &   5.2 &  16.8 &  49.4 \\
  0.179463 &   0.000020 &   131.6 &     7.1 &   0.1016 &   0.0541 &   4.1 &  12.9 &  37.4 \\
  0.182314 &   0.000013 &   318.6 &    10.8 &   0.3090 &   0.0344 &   6.7 &  22.6 &  66.9 \\
  0.201430 &   0.000019 &   159.2 &     7.8 &   0.3960 &   0.0502 &   4.4 &  14.2 &  41.5 \\
  0.241434 &   0.000013 &   329.8 &    11.1 &   0.3809 &   0.0338 &   6.8 &  23.1 &  68.2 \\
  0.243774 &   0.000018 &   169.3 &     8.2 &   0.1483 &   0.0477 &   4.7 &  15.1 &  44.2 \\
  0.257220 &   0.000017 &   178.6 &     8.3 &   0.1921 &   0.0458 &   4.9 &  15.8 &  46.4 \\
  0.258080 &   0.000019 &   140.4 &     7.7 &  -0.1631 &   0.0509 &   4.3 &  13.9 &  40.7 \\
  0.259591 &   0.000020 &   141.1 &     7.2 &   0.4542 &   0.0523 &   4.2 &  13.4 &  38.8 \\
  0.260249 &   0.000013 &   315.6 &    10.9 &   0.0532 &   0.0346 &   6.7 &  22.5 &  66.5 \\
  0.289423 &   0.000020 &   130.6 &     7.1 &  -0.3719 &   0.0536 &   4.1 &  13.0 &  37.7 \\
  0.328324 &   0.000020 &   140.9 &     7.2 &  -0.1802 &   0.0526 &   4.2 &  13.3 &  38.7 \\
  0.328958 &   0.000019 &   147.6 &     7.4 &  -0.4380 &   0.0512 &   4.3 &  13.8 &  40.1 \\
  0.351226 &   0.000013 &   341.1 &    11.4 &   0.3260 &   0.0333 &   6.9 &  23.7 &  70.0 \\
  0.366757 &   0.000020 &   141.7 &     7.3 &   0.4903 &   0.0517 &   4.3 &  13.6 &  39.6 \\
  0.367670 &   0.000009 &   716.7 &    16.4 &   0.1741 &   0.0233 &  10.7 &  37.9 & 111.2 \\
  0.368933 &   0.000017 &   195.4 &     8.6 &  -0.0682 &   0.0441 &   5.1 &  16.5 &  48.4 \\
  0.370250 &   0.000012 &   382.0 &    12.5 &  -0.0796 &   0.0330 &   7.2 &  24.5 &  72.1 \\
  0.371729 &   0.000019 &   144.6 &     7.4 &  -0.4636 &   0.0511 &   4.3 &  13.9 &  40.4 \\
  0.372397 &   0.000021 &   126.7 &     6.8 &  -0.4553 &   0.0548 &   4.0 &  12.6 &  36.3 \\
  0.378262 &   0.000021 &   126.9 &     6.8 &  -0.0795 &   0.0543 &   4.0 &  12.7 &  36.5 \\
  0.386039 &   0.000021 &   135.2 &     7.1 &  -0.0921 &   0.0544 &   4.0 &  12.8 &  37.2 \\
  0.386885 &   0.000013 &   314.2 &    10.7 &   0.3898 &   0.0346 &   6.7 &  22.4 &  66.2 \\
  0.393206 &   0.000019 &   158.1 &     7.6 &   0.1637 &   0.0508 &   4.4 &  14.0 &  40.8 \\
  0.405756 &   0.000013 &   286.5 &    10.3 &  -0.3547 &   0.0354 &   6.6 &  21.6 &  64.1 \\
  0.410438 &   0.000021 &   124.5 &     6.9 &   0.2475 &   0.0543 &   4.0 &  12.8 &  36.8 \\
  0.427893 &   0.000010 &   558.6 &    14.9 &   0.3471 &   0.0274 &   8.8 &  31.0 &  90.9 \\
  0.429327 &   0.000013 &   288.3 &    10.3 &   0.0741 &   0.0357 &   6.5 &  21.4 &  63.5 \\
  0.430994 &   0.000019 &   150.5 &     7.6 &   0.4635 &   0.0507 &   4.4 &  14.0 &  40.9 \\
  0.451546 &   0.000019 &   149.0 &     7.5 &   0.0348 &   0.0513 &   4.3 &  13.8 &  40.3 \\
  0.452014 &   0.000013 &   295.5 &    10.5 &   0.3452 &   0.0350 &   6.6 &  22.0 &  65.2 \\
  0.452940 &   0.000016 &   209.4 &     9.0 &  -0.0821 &   0.0434 &   5.4 &  17.0 &  50.0 \\
  0.453625 &   0.000009 &   740.4 &    16.0 &   0.1075 &   0.0234 &  10.4 &  37.1 & 109.0 \\
  0.454579 &   0.000011 &   528.7 &    14.5 &   0.1139 &   0.0285 &   8.5 &  29.6 &  86.7 \\
  0.455250 &   0.000007 &  1136.1 &    19.0 &   0.2320 &   0.0179 &  14.4 &  51.7 & 151.4 \\
  0.456430 &   0.000019 &   147.8 &     7.8 &   0.1827 &   0.0503 &   4.4 &  14.1 &  41.2 \\
  0.458113 &   0.000013 &   320.5 &    10.6 &   0.0669 &   0.0343 &   6.8 &  22.5 &  66.7 \\
  0.458682 &   0.000020 &   136.5 &     7.2 &  -0.3279 &   0.0532 &   4.1 &  13.1 &  38.1 \\
  0.459223 &   0.000013 &   353.6 &    11.8 &   0.3602 &   0.0336 &   7.0 &  23.7 &  70.1 \\
  0.469579 &   0.000019 &   168.3 &     8.1 &  -0.4609 &   0.0491 &   4.5 &  14.6 &  42.9 \\
  0.472220 &   0.000019 &   148.7 &     7.4 &   0.4996 &   0.0513 &   4.3 &  13.8 &  40.1 \\
  0.477403 &   0.000017 &   187.8 &     8.5 &   0.4358 &   0.0448 &   5.0 &  16.2 &  47.7 \\
  0.486192 &   0.000019 &   162.6 &     8.0 &  -0.2114 &   0.0495 &   4.5 &  14.5 &  42.5 \\
  0.488740 &   0.000020 &   133.8 &     7.0 &  -0.0813 &   0.0538 &   4.1 &  13.0 &  37.5 \\
  0.498027 &   0.000020 &   133.6 &     7.2 &  -0.4488 &   0.0523 &   4.2 &  13.3 &  38.7 \\
  0.498632 &   0.000010 &   527.0 &    14.6 &   0.3366 &   0.0278 &   8.7 &  30.4 &  88.8 \\
  0.500926 &   0.000012 &   449.1 &    13.1 &  -0.1482 &   0.0316 &   7.6 &  25.9 &  76.3 \\
  0.504491 &   0.000012 &   452.8 &    14.1 &  -0.3873 &   0.0311 &   7.8 &  27.0 &  79.1 \\
  0.515126 &   0.000016 &   233.9 &     9.4 &   0.2075 &   0.0420 &   5.6 &  17.8 &  52.5 \\
  0.517303 &   0.000017 &   183.6 &     8.5 &   0.3055 &   0.0461 &   5.0 &  15.8 &  46.5 \\
  0.520767 &   0.000015 &   248.6 &     9.8 &  -0.3921 &   0.0395 &   8.2 &  19.1 &  56.4 \\
  0.529574 &   0.000019 &   148.6 &     7.6 &   0.0260 &   0.0508 &   4.4 &  14.0 &  40.8 \\
  0.530179 &   0.000012 &   442.1 &    13.7 &  -0.2953 &   0.0310 &   7.8 &  26.7 &  78.6 \\
  0.531815 &   0.000016 &   193.5 &     8.9 &   0.1589 &   0.0435 &   5.4 &  16.9 &  49.7 \\
  0.533426 &   0.000006 &  1429.6 &    23.0 &   0.2382 &   0.0158 &  17.0 &  62.8 & 180.4 \\
  0.534037 &   0.000016 &   215.3 &     9.0 &   0.0363 &   0.0436 &   5.4 &  16.9 &  49.8 \\
  0.534769 &   0.000011 &   483.0 &    14.3 &  -0.2111 &   0.0301 &   8.1 &  28.0 &  81.9 \\
  0.535448 &   0.000015 &   216.7 &     9.7 &  -0.2485 &   0.0403 &   5.8 &  18.6 &  55.1 \\
  0.536294 &   0.000009 &   755.4 &    16.7 &   0.1368 &   0.0236 &  10.7 &  37.7 & 110.7 \\
  0.536881 &   0.000012 &   516.4 &    13.9 &  -0.1059 &   0.0311 &   7.4 &  26.8 &  78.7 \\
  0.537435 &   0.000007 &  1083.8 &    18.2 &   0.1906 &   0.0189 &  13.4 &  48.3 & 141.0 \\
  0.540373 &   0.000017 &   197.2 &     8.7 &  -0.1831 &   0.0438 &   5.3 &  16.7 &  49.0 \\
  0.548936 &   0.000013 &   346.4 &    11.3 &   0.0673 &   0.0333 &   7.1 &  23.5 &  69.6 \\
  0.549494 &   0.000007 &  1042.2 &    19.9 &   0.0965 &   0.0175 &  14.7 &  53.5 & 155.6 \\
  0.550267 &   0.000013 &   375.5 &    11.9 &   0.0187 &   0.0335 &   7.2 &  23.9 &  70.4 \\
  0.551086 &   0.000007 &   992.5 &    18.6 &  -0.2317 &   0.0188 &  13.6 &  49.0 & 143.3 \\
  0.551610 &   0.000015 &   260.2 &     9.8 &  -0.0390 &   0.0386 &   6.2 &  19.5 &  57.6 \\
  0.552608 &   0.000002 &  8738.6 &    57.0 &   0.3018 &   0.0062 &  51.5 & 200.9 & 518.8 \\
  0.553582 &   0.000012 &   334.5 &    12.6 &  -0.1969 &   0.0327 &   7.4 &  24.8 &  72.9 \\
  0.554482 &   0.000013 &   294.6 &    10.4 &  -0.4830 &   0.0351 &   6.8 &  21.8 &  64.6 \\
  0.555084 &   0.000006 &  1424.9 &    21.6 &   0.2724 &   0.0159 &  16.6 &  60.1 & 173.6 \\
  0.555670 &   0.000003 &  5106.2 &    42.8 &  -0.4922 &   0.0085 &  36.1 & 140.7 & 375.4 \\
  0.556445 &   0.000013 &   330.6 &    11.0 &   0.2613 &   0.0343 &   6.9 &  22.7 &  67.1 \\
  0.557035 &   0.000021 &   117.2 &     6.5 &   0.3264 &   0.0554 &   4.1 &  12.3 &  35.4 \\
  0.568206 &   0.000013 &   335.5 &    11.7 &   0.2336 &   0.0336 &   7.1 &  23.6 &  69.8 \\
  0.569070 &   0.000002 & 14746.3 &    91.1 &  -0.2775 &   0.0063 &  53.7 & 213.1 & 534.1 \\
  0.570719 &   0.000004 &  2091.3 &    23.7 &  -0.2568 &   0.0116 &  23.1 &  85.8 & 246.1 \\
  0.571389 &   0.000012 &   397.7 &    12.9 &   0.0010 &   0.0330 &   7.3 &  24.6 &  72.4 \\
  0.571964 &   0.000004 &  3017.3 &    28.7 &  -0.4514 &   0.0098 &  28.9 & 109.2 & 305.3 \\
  0.572072 &   0.000021 &   121.1 &     6.6 &  -0.1203 &   0.0552 &   4.1 &  12.4 &  35.7 \\
  0.572793 &   0.000013 &   337.7 &    11.1 &   0.1697 &   0.0343 &   6.9 &  22.7 &  67.0 \\
  0.573571 &   0.000009 &   677.0 &    15.6 &  -0.1578 &   0.0235 &  10.4 &  36.7 & 107.5 \\
  0.574156 &   0.000003 &  3963.9 &    34.6 &   0.2836 &   0.0084 &  35.5 & 136.2 & 371.9 \\
  0.574908 &   0.000009 &   703.5 &    15.8 &  -0.1305 &   0.0241 &  10.2 &  35.8 & 105.0 \\
  0.575513 &   0.000020 &   128.7 &     7.0 &  -0.0768 &   0.0542 &   4.2 &  12.8 &  36.9 \\
  0.585671 &   0.000015 &   249.4 &     9.9 &   0.2886 &   0.0389 &   6.1 &  19.4 &  57.5 \\
  0.586103 &   0.000020 &   129.6 &     7.0 &  -0.3528 &   0.0542 &   4.2 &  12.9 &  37.2 \\
  0.589465 &   0.000019 &   157.0 &     7.9 &  -0.3170 &   0.0504 &   4.5 &  14.2 &  41.4 \\
  0.590231 &   0.000012 &   386.2 &    12.3 &  -0.4284 &   0.0329 &   7.3 &  24.4 &  71.9 \\
  0.590939 &   0.000012 &   432.3 &    13.6 &   0.2294 &   0.0315 &   7.7 &  26.2 &  77.0 \\
  0.592033 &   0.000016 &   219.0 &     9.4 &   0.3679 &   0.0418 &   5.7 &  17.8 &  52.5 \\
  0.593598 &   0.000013 &   385.2 &    12.2 &   0.3392 &   0.0331 &   7.3 &  24.2 &  71.4 \\
  0.594659 &   0.000018 &   183.1 &     8.4 &   0.4779 &   0.0466 &   5.1 &  15.6 &  45.8 \\
  0.596196 &   0.000012 &   434.1 &    13.1 &  -0.0740 &   0.0322 &   7.6 &  25.4 &  74.6 \\
  0.597990 &   0.000019 &   142.4 &     7.5 &  -0.0745 &   0.0511 &   4.5 &  13.8 &  40.3 \\
  0.612314 &   0.000015 &   241.2 &     9.6 &  -0.3986 &   0.0409 &   5.7 &  18.3 &  54.0 \\
  0.612913 &   0.000004 &  2708.2 &    26.8 &   0.1584 &   0.0102 &  26.6 & 101.6 & 286.8 \\
  0.613674 &   0.000016 &   214.3 &     9.5 &   0.3604 &   0.0417 &   5.6 &  17.9 &  52.8 \\
  0.614552 &   0.000013 &   376.2 &    12.1 &   0.1459 &   0.0333 &   7.1 &  24.0 &  70.9 \\
  0.615472 &   0.000006 &  1342.3 &    20.9 &  -0.2840 &   0.0157 &  16.3 &  60.5 & 174.7 \\
  0.615855 &   0.000016 &   220.6 &     9.1 &   0.0450 &   0.0429 &   5.4 &  17.3 &  50.9 \\
  0.616586 &   0.000020 &   142.0 &     7.3 &   0.0739 &   0.0522 &   4.3 &  13.4 &  39.1 \\
  0.617935 &   0.000020 &   129.2 &     7.1 &   0.1375 &   0.0530 &   4.3 &  13.2 &  38.1 \\
  0.618628 &   0.000004 &  2437.8 &    25.0 &  -0.2197 &   0.0103 &  25.4 &  97.1 & 275.9 \\
  0.619208 &   0.000016 &   228.8 &     9.2 &  -0.4434 &   0.0419 &   5.6 &  17.7 &  52.3 \\
  0.638671 &   0.000016 &   217.8 &     9.3 &   0.1563 &   0.0416 &   5.5 &  17.8 &  52.6 \\
  0.665835 &   0.000019 &   150.5 &     7.6 &  -0.3766 &   0.0508 &   4.4 &  14.0 &  40.8 \\
  0.667090 &   0.000019 &   172.5 &     8.0 &  -0.4656 &   0.0505 &   4.5 &  14.2 &  41.6 \\
  0.668309 &   0.000013 &   342.7 &    11.2 &   0.1900 &   0.0339 &   6.7 &  23.2 &  68.5 \\
  0.669395 &   0.000021 &   132.3 &     7.0 &   0.2756 &   0.0544 &   4.2 &  12.7 &  36.8 \\
  0.670600 &   0.000019 &   156.7 &     7.5 &   0.0633 &   0.0512 &   4.4 &  13.8 &  40.4 \\
  0.674313 &   0.000016 &   211.4 &     8.9 &  -0.3565 &   0.0433 &   5.3 &  17.0 &  50.0 \\
  0.675489 &   0.000014 &   281.0 &    10.2 &   0.4530 &   0.0358 &   6.4 &  21.3 &  63.1 \\
  0.676374 &   0.000021 &   129.8 &     7.0 &  -0.2114 &   0.0544 &   5.2 &  12.8 &  37.1 \\
  0.698686 &   0.000016 &   238.4 &     9.6 &   0.0512 &   0.0415 &   5.5 &  18.0 &  53.3 \\
  0.703651 &   0.000013 &   385.2 &    12.3 &  -0.0731 &   0.0333 &   6.9 &  24.2 &  71.2 \\
  0.731048 &   0.000019 &   157.9 &     7.7 &  -0.2928 &   0.0507 &   4.4 &  14.0 &  40.8 \\
  0.731687 &   0.000006 &  1399.0 &    22.3 &  -0.3638 &   0.0154 &  16.6 &  63.4 & 182.4 \\
  0.732383 &   0.000003 &  3860.3 &    30.9 &  -0.0542 &   0.0079 &  36.1 & 139.9 & 386.2 \\
  0.733017 &   0.000007 &  1081.1 &    19.4 &   0.3086 &   0.0183 &  13.4 &  51.0 & 148.7 \\
  0.733801 &   0.000013 &   365.2 &    11.8 &   0.1260 &   0.0338 &   6.8 &  23.6 &  69.7 \\
  0.734708 &   0.000003 &  9640.6 &    67.5 &  -0.1626 &   0.0070 &  48.0 & 190.9 & 487.8 \\
  0.735360 &   0.000012 &   403.8 &    13.3 &   0.4714 &   0.0319 &   7.3 &  25.8 &  75.9 \\
  0.736156 &   0.000016 &   225.7 &     9.5 &  -0.3664 &   0.0412 &   5.5 &  18.2 &  53.7 \\
  0.736776 &   0.000008 &   810.5 &    16.9 &  -0.1776 &   0.0214 &  11.2 &  41.7 & 122.4 \\
  0.737481 &   0.000002 & 12307.1 &    77.4 &  -0.2389 &   0.0063 &  52.6 & 210.4 & 529.7 \\
  0.738047 &   0.000016 &   204.7 &     8.8 &   0.1311 &   0.0436 &   5.2 &  16.7 &  49.2 \\
  0.738824 &   0.000020 &   140.5 &     7.2 &   0.1121 &   0.0525 &   4.3 &  13.3 &  38.7 \\
  0.745159 &   0.000027 &    54.3 &     4.1 &   0.2114 &   0.0725 &   4.3 &   8.5 &  23.2 \\
  0.747771 &   0.000021 &   131.8 &     6.7 &   0.1060 &   0.0548 &   4.1 &  12.5 &  36.1 \\
  0.751222 &   0.000016 &   205.1 &     9.0 &   0.1810 &   0.0433 &   5.3 &  17.0 &  50.2 \\
  0.753934 &   0.000020 &   130.6 &     6.8 &  -0.3673 &   0.0541 &   4.2 &  12.7 &  36.7 \\
  0.770238 &   0.000012 &   452.4 &    14.0 &  -0.2974 &   0.0308 &   7.6 &  27.1 &  79.4 \\
  0.772399 &   0.000015 &   258.6 &    10.0 &  -0.2066 &   0.0389 &   5.9 &  19.5 &  57.8 \\
  0.775107 &   0.000007 &   901.2 &    17.8 &  -0.0435 &   0.0194 &  12.5 &  46.6 & 136.0 \\
  0.775628 &   0.000020 &   121.4 &     6.7 &  -0.2063 &   0.0538 &   4.2 &  12.8 &  36.8 \\
  0.809285 &   0.000019 &   163.3 &     8.1 &   0.2412 &   0.0494 &   4.6 &  14.5 &  42.5 \\
  0.809859 &   0.000010 &   561.6 &    15.1 &   0.2744 &   0.0276 &   8.8 &  30.9 &  90.6 \\
  0.810634 &   0.000003 &  4960.1 &    38.2 &   0.0099 &   0.0077 &  37.4 & 148.7 & 400.6 \\
  0.811231 &   0.000016 &   204.4 &     8.7 &   0.0770 &   0.0437 &   5.2 &  16.7 &  49.2 \\
  0.812676 &   0.000018 &   186.8 &     8.4 &  -0.1171 &   0.0466 &   4.9 &  15.6 &  45.9 \\
  0.812940 &   0.000003 &  5601.8 &    51.8 &   0.4581 &   0.0085 &  37.2 & 149.5 & 392.4 \\
  0.813394 &   0.000013 &   300.2 &    10.4 &  -0.2769 &   0.0352 &   6.4 &  21.8 &  64.4 \\
  0.814128 &   0.000016 &   205.2 &     9.3 &  -0.4518 &   0.0417 &   5.5 &  17.8 &  52.7 \\
  0.814689 &   0.000012 &   442.1 &    13.8 &  -0.2008 &   0.0308 &   7.5 &  27.0 &  79.3 \\
  0.815392 &   0.000003 &  5514.4 &    47.2 &   0.1142 &   0.0087 &  35.7 & 143.9 & 379.7 \\
  0.816091 &   0.000008 &   863.3 &    17.5 &  -0.4043 &   0.0205 &  11.5 &  43.7 & 128.1 \\
  0.816720 &   0.000016 &   213.1 &     9.3 &  -0.4268 &   0.0418 &   5.5 &  17.8 &  52.4 \\
  0.854235 &   0.000012 &   415.3 &    13.2 &   0.4031 &   0.0319 &   7.3 &  25.7 &  75.6 \\
  0.856351 &   0.000010 &   510.2 &    14.8 &  -0.3446 &   0.0270 &   8.7 &  31.5 &  92.1 \\
  0.858747 &   0.000012 &   426.3 &    13.4 &   0.0143 &   0.0319 &   7.3 &  25.9 &  76.1 \\
  0.899948 &   0.000015 &   256.4 &     9.9 &   0.3139 &   0.0391 &   5.9 &  19.3 &  57.0 \\
  0.904512 &   0.000019 &   155.2 &     7.8 &  -0.0632 &   0.0502 &   4.7 &  14.2 &  41.5 \\
  0.951900 &   0.000013 &   306.4 &    10.6 &  -0.0379 &   0.0343 &   6.9 &  22.5 &  66.7 \\
  0.954107 &   0.000010 &   584.5 &    15.4 &   0.3980 &   0.0264 &   9.4 &  32.6 &  95.6 \\
  0.956347 &   0.000013 &   342.1 &    10.9 &  -0.2927 &   0.0343 &   7.1 &  22.6 &  66.9 \\
  0.992043 &   0.000019 &   137.9 &     7.3 &   0.1470 &   0.0516 &   4.8 &  13.6 &  39.7 \\
  1.007832 &   0.000023 &    94.9 &     5.7 &  -0.2763 &   0.0602 &   4.0 &  11.0 &  31.3 \\
  1.011053 &   0.000019 &   150.6 &     7.6 &  -0.3821 &   0.0510 &   4.9 &  13.9 &  40.6 \\
  1.011792 &   0.000014 &   267.9 &    10.1 &  -0.4959 &   0.0376 &   6.5 &  20.3 &  60.0 \\
  1.013415 &   0.000012 &   376.9 &    12.7 &  -0.0778 &   0.0328 &   7.7 &  24.7 &  72.6 \\
  1.013935 &   0.000023 &    92.5 &     5.5 &   0.3177 &   0.0608 &   4.0 &  10.8 &  30.7 \\
  1.014504 &   0.000016 &   200.3 &     8.7 &   0.0298 &   0.0435 &   5.8 &  16.7 &  49.2 \\
  1.015822 &   0.000008 &   804.3 &    17.2 &  -0.2938 &   0.0214 &  12.6 &  41.8 & 122.6 \\
  1.016804 &   0.000022 &    98.0 &     6.1 &  -0.2332 &   0.0589 &   4.2 &  11.4 &  32.6 \\
  1.029620 &   0.000021 &   124.9 &     6.7 &   0.1304 &   0.0545 &   4.6 &  12.6 &  36.3 \\
  1.032802 &   0.000019 &   158.1 &     7.9 &   0.1338 &   0.0500 &   5.1 &  14.3 &  41.8 \\
  1.055115 &   0.000021 &   126.9 &     6.8 &   0.4180 &   0.0546 &   4.7 &  12.6 &  36.3 \\
  1.061753 &   0.000023 &    82.7 &     5.4 &  -0.2403 &   0.0621 &   4.1 &  10.5 &  29.7 \\
  1.086757 &   0.000022 &   101.0 &     5.9 &  -0.3135 &   0.0593 &   4.4 &  11.3 &  32.1 \\
  1.098815 &   0.000019 &   154.1 &     7.7 &  -0.0952 &   0.0508 &   5.4 &  14.0 &  40.9 \\
  1.099532 &   0.000016 &   205.8 &     9.1 &  -0.4633 &   0.0429 &   6.4 &  17.2 &  50.8 \\
  1.101844 &   0.000023 &    92.8 &     5.6 &   0.1533 &   0.0601 &   4.4 &  11.0 &  31.3 \\
  1.102313 &   0.000010 &   579.5 &    15.2 &   0.3949 &   0.0260 &  11.9 &  32.9 &  96.3 \\
  1.103526 &   0.000019 &   163.5 &     7.9 &   0.2864 &   0.0506 &   5.4 &  14.1 &  41.4 \\
  1.105236 &   0.000012 &   366.3 &    13.0 &  -0.2616 &   0.0330 &   9.0 &  24.7 &  72.5 \\
  1.106130 &   0.000018 &   181.1 &     8.4 &   0.3444 &   0.0464 &   6.0 &  15.6 &  46.0 \\
  1.108261 &   0.000015 &   265.0 &    10.0 &   0.1326 &   0.0386 &   7.1 &  19.7 &  58.3 \\
  1.112839 &   0.000019 &   146.6 &     7.5 &   0.1254 &   0.0510 &   5.4 &  13.9 &  40.5 \\
  1.121714 &   0.000013 &   286.7 &    10.1 &  -0.1977 &   0.0356 &   7.9 &  21.4 &  63.3 \\
  1.123448 &   0.000022 &   100.6 &     5.9 &   0.0843 &   0.0586 &   4.6 &  11.4 &  32.4 \\
  1.124155 &   0.000018 &   172.8 &     8.2 &   0.1848 &   0.0470 &   5.9 &  15.3 &  45.0 \\
  1.124824 &   0.000013 &   335.6 &    11.5 &   0.2777 &   0.0335 &   8.9 &  23.5 &  69.5 \\
  1.129776 &   0.000024 &    82.7 &     5.2 &  -0.3852 &   0.0638 &   4.2 &  10.2 &  28.6 \\
  1.143149 &   0.000013 &   349.6 &    11.4 &   0.1868 &   0.0334 &   9.1 &  23.6 &  69.9 \\
  1.165411 &   0.000019 &   154.4 &     7.8 &   0.2916 &   0.0502 &   5.7 &  14.2 &  41.5 \\
  1.168373 &   0.000016 &   208.8 &     8.6 &  -0.0624 &   0.0433 &   6.7 &  16.9 &  49.5 \\
  1.170061 &   0.000023 &    92.3 &     5.5 &   0.1173 &   0.0603 &   4.5 &  10.9 &  30.9 \\
  1.171198 &   0.000019 &   158.9 &     8.0 &  -0.1237 &   0.0505 &   5.7 &  14.2 &  41.5 \\
  1.172187 &   0.000023 &    92.4 &     5.4 &   0.2960 &   0.0607 &   4.5 &  10.8 &  30.5 \\
  1.173801 &   0.000025 &    77.5 &     4.9 &  -0.1136 &   0.0668 &   4.0 &   9.6 &  26.7 \\
  1.174400 &   0.000021 &   128.5 &     7.0 &   0.0819 &   0.0545 &   5.2 &  12.8 &  36.9 \\
  1.186909 &   0.000023 &    91.5 &     5.7 &  -0.2643 &   0.0599 &   4.7 &  11.0 &  31.3 \\
  1.187774 &   0.000016 &   222.0 &     9.2 &  -0.4579 &   0.0417 &   7.1 &  17.7 &  52.2 \\
  1.190025 &   0.000024 &    82.5 &     5.2 &   0.1559 &   0.0639 &   4.3 &  10.2 &  28.5 \\
  1.206430 &   0.000017 &   198.8 &     8.6 &  -0.3525 &   0.0442 &   6.9 &  16.5 &  48.4 \\
  1.206955 &   0.000026 &    71.9 &     4.8 &  -0.0201 &   0.0676 &   4.1 &   9.4 &  26.2 \\
  1.215183 &   0.000022 &   101.6 &     5.9 &  -0.0744 &   0.0586 &   4.9 &  11.4 &  32.5 \\
  1.244516 &   0.000030 &    43.2 &     3.4 &   0.2773 &   0.0796 &   4.2 &   7.5 &  19.6 \\
  1.262688 &   0.000015 &   247.5 &     9.7 &  -0.3052 &   0.0393 &   8.2 &  19.1 &  56.6 \\
  1.264449 &   0.000019 &   145.8 &     7.4 &  -0.1641 &   0.0510 &   6.1 &  13.8 &  40.3 \\
  1.265081 &   0.000021 &   126.6 &     6.7 &  -0.3899 &   0.0546 &   5.5 &  12.5 &  36.1 \\
  1.266994 &   0.000019 &   160.5 &     7.7 &  -0.0945 &   0.0509 &   6.2 &  13.9 &  40.7 \\
  1.268016 &   0.000019 &   183.3 &     7.9 &   0.1007 &   0.0506 &   4.5 &  14.1 &  41.3 \\
  1.269341 &   0.000025 &    78.7 &     5.1 &  -0.0103 &   0.0664 &   4.3 &   9.7 &  27.3 \\
  1.271013 &   0.000018 &   175.8 &     8.2 &   0.1826 &   0.0469 &   6.7 &  15.4 &  45.1 \\
  1.272301 &   0.000024 &    91.7 &     5.4 &   0.0876 &   0.0626 &   4.7 &  10.4 &  29.4 \\
  1.275151 &   0.000021 &   116.7 &     6.5 &   0.3269 &   0.0556 &   5.5 &  12.3 &  35.2 \\
  1.284692 &   0.000022 &   124.2 &     6.1 &   0.2840 &   0.0580 &   5.1 &  11.6 &  33.2 \\
  1.285579 &   0.000022 &   101.0 &     5.9 &   0.4668 &   0.0588 &   5.1 &  11.4 &  32.5 \\
  1.287386 &   0.000009 &   698.6 &    16.2 &   0.1946 &   0.0233 &  16.6 &  37.7 & 110.8 \\
  1.289019 &   0.000025 &    75.6 &     4.9 &   0.2470 &   0.0670 &   4.4 &   9.6 &  26.7 \\
  1.289951 &   0.000012 &   364.3 &    12.8 &  -0.0526 &   0.0330 &  11.0 &  24.6 &  72.2 \\
  1.292561 &   0.000023 &    91.9 &     5.5 &  -0.1836 &   0.0604 &   4.9 &  10.9 &  30.8 \\
  1.293147 &   0.000013 &   335.5 &    11.6 &   0.0728 &   0.0337 &  10.4 &  23.4 &  69.4 \\
  1.301501 &   0.000022 &   111.7 &     6.2 &   0.4033 &   0.0575 &   5.3 &  11.8 &  33.7 \\
  1.303764 &   0.000016 &   212.6 &     9.1 &   0.3601 &   0.0431 &   7.7 &  17.2 &  50.6 \\
  1.305327 &   0.000020 &   142.1 &     7.3 &  -0.2272 &   0.0518 &   6.2 &  13.6 &  39.4 \\
  1.306578 &   0.000006 &  1221.4 &    20.4 &  -0.4726 &   0.0165 &  26.7 &  57.2 & 165.7 \\
  1.308670 &   0.000026 &    78.8 &     4.9 &  -0.0563 &   0.0677 &   4.4 &   9.5 &  26.3 \\
  1.309476 &   0.000025 &    73.7 &     4.9 &   0.0946 &   0.0669 &   4.4 &   9.5 &  26.5 \\
  1.311662 &   0.000021 &   127.2 &     6.9 &  -0.2071 &   0.0543 &   4.0 &  12.7 &  36.7 \\
  1.338724 &   0.000025 &    74.1 &     4.8 &  -0.0149 &   0.0673 &   4.5 &   9.5 &  26.4 \\
  1.344107 &   0.000023 &    94.7 &     5.8 &  -0.1641 &   0.0601 &   5.1 &  11.1 &  31.4 \\
  1.345288 &   0.000026 &    62.8 &     4.5 &  -0.0809 &   0.0698 &   4.2 &   9.1 &  24.9 \\
  1.346759 &   0.000027 &    61.7 &     4.4 &   0.4517 &   0.0705 &   4.2 &   8.9 &  24.3 \\
  1.347807 &   0.000021 &   113.9 &     6.4 &  -0.4380 &   0.0567 &   5.5 &  12.0 &  34.4 \\
  1.353563 &   0.000023 &    96.2 &     5.9 &   0.2495 &   0.0596 &   5.3 &  11.2 &  31.9 \\
  1.356076 &   0.000018 &   168.3 &     8.1 &   0.3519 &   0.0484 &   6.9 &  14.8 &  43.5 \\
  1.363303 &   0.000020 &   118.6 &     6.9 &   0.0580 &   0.0542 &   6.0 &  12.8 &  36.9 \\
  1.365498 &   0.000012 &   386.0 &    12.4 &  -0.1236 &   0.0329 &  11.9 &  24.6 &  72.4 \\
  1.366224 &   0.000022 &   107.3 &     6.2 &  -0.3383 &   0.0577 &   5.5 &  11.7 &  33.5 \\
  1.368282 &   0.000018 &   171.2 &     8.2 &   0.0246 &   0.0473 &   7.1 &  15.2 &  44.7 \\
  1.371044 &   0.000025 &    71.9 &     4.8 &   0.4470 &   0.0675 &   4.5 &   9.5 &  26.3 \\
  1.379722 &   0.000022 &   109.7 &     6.3 &   0.4837 &   0.0570 &   5.6 &  11.9 &  34.1 \\
  1.381988 &   0.000019 &   142.4 &     7.3 &  -0.0409 &   0.0516 &   6.6 &  13.6 &  39.6 \\
  1.384455 &   0.000013 &   353.7 &    12.0 &  -0.1379 &   0.0335 &  11.6 &  23.9 &  70.5 \\
  1.385026 &   0.000023 &    94.4 &     5.6 &   0.1085 &   0.0602 &   5.2 &  11.0 &  31.1 \\
  1.387190 &   0.000028 &    54.6 &     4.0 &  -0.4664 &   0.0740 &   4.1 &   8.4 &  22.6 \\
  1.403786 &   0.000025 &    76.2 &     5.0 &   0.4602 &   0.0667 &   4.7 &   9.6 &  26.9 \\
  1.408874 &   0.000021 &   131.4 &     7.1 &   0.3578 &   0.0543 &   6.2 &  12.9 &  37.3 \\
  1.409610 &   0.000026 &    68.9 &     4.7 &   0.0633 &   0.0682 &   4.6 &   9.3 &  25.8 \\
  1.410319 &   0.000021 &   124.1 &     6.6 &  -0.0889 &   0.0553 &   6.3 &  12.4 &  35.6 \\
  1.412664 &   0.000017 &   183.3 &     8.5 &   0.2839 &   0.0461 &   5.0 &  15.8 &  46.4 \\
  1.413752 &   0.000021 &   122.4 &     6.6 &   0.4601 &   0.0552 &   6.1 &  12.4 &  35.6 \\
  1.423569 &   0.000028 &    51.4 &     3.9 &  -0.1370 &   0.0742 &   4.1 &   8.3 &  22.2 \\
  1.428287 &   0.000019 &   161.5 &     8.0 &   0.3291 &   0.0505 &   6.9 &  14.1 &  41.4 \\
  1.429292 &   0.000028 &    55.9 &     4.1 &  -0.2060 &   0.0734 &   4.2 &   8.4 &  22.8 \\
  1.431478 &   0.000027 &    58.8 &     4.1 &   0.1117 &   0.0723 &   4.3 &   8.6 &  23.3 \\
  1.467143 &   0.000022 &   105.4 &     6.0 &  -0.0328 &   0.0593 &   5.7 &  11.3 &  32.3 \\
  1.469356 &   0.000021 &   119.0 &     6.6 &  -0.1816 &   0.0550 &   6.4 &  12.5 &  35.8 \\
  1.470170 &   0.000016 &   198.8 &     8.8 &   0.4638 &   0.0437 &   8.4 &  16.8 &  49.5 \\
  1.472166 &   0.000015 &   231.6 &     9.6 &  -0.1501 &   0.0408 &   9.4 &  18.4 &  54.5 \\
  1.475014 &   0.000019 &   141.8 &     7.4 &  -0.2947 &   0.0515 &   4.3 &  13.7 &  39.9 \\
  1.482560 &   0.000021 &   120.5 &     6.7 &  -0.2010 &   0.0546 &   4.6 &  12.6 &  36.3 \\
  1.483677 &   0.000029 &    47.4 &     3.6 &   0.2608 &   0.0772 &   4.0 &   7.8 &  20.7 \\
  1.486466 &   0.000025 &    77.2 &     5.1 &   0.1932 &   0.0659 &   5.1 &   9.8 &  27.4 \\
  1.487377 &   0.000029 &    50.3 &     3.8 &  -0.0481 &   0.0758 &   4.2 &   8.1 &  21.6 \\
  1.547329 &   0.000021 &   127.5 &     6.9 &  -0.2910 &   0.0544 &   7.0 &  12.7 &  36.8 \\
  1.547877 &   0.000011 &   474.5 &    14.2 &   0.0768 &   0.0300 &  17.0 &  28.0 &  81.9 \\
  1.548476 &   0.000023 &    97.5 &     5.8 &   0.4573 &   0.0597 &   6.1 &  11.2 &  31.8 \\
  1.554951 &   0.000030 &    40.9 &     3.3 &   0.0430 &   0.0805 &   4.0 &   7.4 &  19.1 \\
  1.560704 &   0.000029 &    48.2 &     3.7 &  -0.3755 &   0.0762 &   4.3 &   8.0 &  21.1 \\
  1.625918 &   0.000023 &    95.0 &     5.7 &   0.1163 &   0.0605 &   6.8 &  11.0 &  31.2 \\
  1.654821 &   0.000026 &    70.3 &     4.7 &  -0.1076 &   0.0679 &   6.0 &   9.4 &  26.0 \\
  1.657728 &   0.000019 &   160.2 &     8.0 &   0.2437 &   0.0501 &   9.2 &  14.3 &  42.0 \\
  1.689045 &   0.000032 &    35.3 &     3.0 &   0.1040 &   0.0841 &   4.3 &   6.9 &  17.6 \\
  1.744530 &   0.000029 &    47.1 &     3.7 &   0.3034 &   0.0760 &   5.3 &   8.0 &  21.2 \\
  1.745275 &   0.000027 &    62.4 &     4.3 &   0.1727 &   0.0705 &   6.2 &   8.9 &  24.3 \\
  1.748169 &   0.000029 &    45.8 &     3.6 &  -0.3588 &   0.0776 &   5.2 &   7.8 &  20.5 \\
  1.748761 &   0.000032 &    36.0 &     3.0 &   0.4024 &   0.0841 &   4.6 &   6.9 &  17.6 \\
  1.750237 &   0.000027 &    60.3 &     4.3 &   0.0429 &   0.0710 &   6.1 &   8.8 &  24.0 \\
  1.751529 &   0.000025 &    74.8 &     5.0 &   0.3985 &   0.0672 &   7.0 &   9.6 &  26.8 \\
  1.756882 &   0.000029 &    49.7 &     3.7 &  -0.3251 &   0.0758 &   5.5 &   8.0 &  21.4 \\
  1.836973 &   0.000031 &    36.5 &     3.0 &   0.3842 &   0.0833 &   5.1 &   7.0 &  17.8 \\
  1.840034 &   0.000027 &    59.3 &     4.2 &  -0.0280 &   0.0723 &   6.8 &   8.6 &  23.5 \\
  1.842789 &   0.000028 &    55.6 &     4.1 &   0.1741 &   0.0734 &   6.6 &   8.4 &  22.8 \\
  1.859246 &   0.000033 &    30.8 &     2.7 &  -0.4372 &   0.0881 &   4.8 &   6.5 &  16.0 \\
  1.862310 &   0.000031 &    37.3 &     3.1 &  -0.2389 &   0.0829 &   5.4 &   7.1 &  18.1 \\
  1.873265 &   0.000024 &    80.9 &     5.2 &   0.0085 &   0.0641 &   9.2 &  10.1 &  28.4 \\
  1.877064 &   0.000035 &    27.7 &     2.5 &  -0.3646 &   0.0914 &   4.5 &   6.2 &  14.9 \\
  1.915540 &   0.000035 &    26.6 &     2.4 &  -0.2925 &   0.0918 &   4.7 &   6.1 &  14.8 \\
  1.978247 &   0.000036 &    20.5 &     2.0 &   0.1552 &   0.0964 &   4.3 &   5.7 &  12.7 \\
  1.981865 &   0.000034 &    29.4 &     2.6 &  -0.0430 &   0.0892 &   5.4 &   6.4 &  15.7 \\
  1.986711 &   0.000036 &    22.1 &     2.1 &   0.3570 &   0.0947 &   4.4 &   5.8 &  13.3 \\
  2.000206 &   0.000034 &    28.4 &     2.6 &   0.4966 &   0.0902 &   5.5 &   6.3 &  15.4 \\
  2.041291 &   0.000031 &    38.0 &     3.2 &  -0.3941 &   0.0829 &   6.9 &   7.1 &  18.3 \\
  2.044121 &   0.000030 &    42.7 &     3.4 &   0.4643 &   0.0796 &   7.4 &   7.5 &  19.5 \\
  2.103660 &   0.000039 &    16.3 &     1.7 &   0.3335 &   0.1042 &   4.1 &   5.1 &  10.9 \\
  2.204819 &   0.000037 &    19.0 &     1.9 &  -0.4009 &   0.0985 &   5.1 &   5.5 &  12.2 \\
  2.209813 &   0.000038 &    17.9 &     1.8 &   0.4496 &   0.1002 &   4.9 &   5.4 &  11.7 \\
  2.226761 &   0.000040 &    15.9 &     1.7 &   0.0299 &   0.1052 &   4.5 &   5.1 &  10.8 \\
  2.282531 &   0.000035 &    24.5 &     2.2 &   0.2270 &   0.0927 &   6.4 &   6.0 &  13.9 \\
  2.285430 &   0.000035 &    23.6 &     2.2 &   0.0339 &   0.0928 &   4.3 &   6.0 &  13.9 \\
  2.301520 &   0.000043 &    12.5 &     1.4 &   0.2681 &   0.1146 &   4.1 &   4.6 &   9.2 \\
  2.322422 &   0.000041 &    14.0 &     1.5 &   0.4634 &   0.1087 &   4.4 &   4.9 &   9.9 \\
  2.325772 &   0.000040 &    15.1 &     1.6 &  -0.2756 &   0.1066 &   4.7 &   5.0 &  10.4 \\
  2.360716 &   0.000043 &    12.3 &     1.4 &  -0.0608 &   0.1151 &   4.2 &   4.6 &   9.0 \\
  2.473027 &   0.000038 &    17.5 &     1.8 &   0.3830 &   0.1005 &   5.9 &   5.4 &  11.5 \\
  2.482706 &   0.000041 &    13.8 &     1.5 &   0.0732 &   0.1095 &   5.0 &   4.9 &   9.8 \\
  2.486397 &   0.000042 &    13.3 &     1.5 &   0.1876 &   0.1119 &   4.9 &   4.7 &   9.5 \\
  2.555695 &   0.000036 &    21.6 &     2.1 &   0.0841 &   0.0952 &   7.6 &   5.8 &  13.1 \\
  2.717759 &   0.000043 &    12.6 &     1.4 &   0.1519 &   0.1147 &   5.4 &   4.6 &   9.1 \\
  2.719520 &   0.000049 &     9.1 &     1.2 &  -0.3600 &   0.1306 &   4.2 &   4.1 &   7.3 \\
  2.721429 &   0.000042 &    13.8 &     1.5 &   0.1734 &   0.1102 &   5.9 &   4.8 &   9.8 \\
  2.778804 &   0.000045 &    10.8 &     1.3 &   0.1275 &   0.1204 &   4.9 &   4.4 &   8.3 \\
  2.798246 &   0.000050 &     8.8 &     1.2 &   0.1628 &   0.1335 &   4.1 &   4.0 &   7.0 \\
  2.815631 &   0.000046 &    10.4 &     1.3 &  -0.1842 &   0.1229 &   4.9 &   4.3 &   8.1 \\
  2.828293 &   0.000052 &     8.0 &     1.1 &  -0.4090 &   0.1375 &   4.1 &   3.9 &   6.7 \\
  2.931545 &   0.000049 &     9.1 &     1.2 &   0.2674 &   0.1302 &   4.6 &   4.1 &   7.3 \\
  2.935118 &   0.000047 &    10.4 &     1.3 &  -0.4157 &   0.1232 &   5.1 &   4.3 &   8.0 \\
  2.987599 &   0.000039 &    16.3 &     1.7 &  -0.0918 &   0.1038 &   7.9 &   5.2 &  11.0 \\
  3.042066 &   0.000043 &    12.5 &     1.4 &  -0.0289 &   0.1148 &   6.5 &   4.7 &   9.1 \\
  3.389617 &   0.000044 &    11.9 &     1.4 &   0.4923 &   0.1173 &   7.4 &   4.9 &   8.8 \\
  3.602889 &   0.000049 &     9.3 &     1.2 &  -0.1124 &   0.1293 &   6.4 &   4.7 &   7.4 \\
  3.605013 &   0.000059 &     6.4 &     1.0 &  -0.2848 &   0.1569 &   4.5 &   3.9 &   5.6 \\
  3.609567 &   0.000052 &     7.8 &     1.1 &  -0.4552 &   0.1384 &   5.4 &   4.3 &   6.5 \\
  3.611211 &   0.000054 &     8.2 &     1.1 &  -0.3054 &   0.1437 &   5.2 &   4.2 &   6.3 \\
  3.625541 &   0.000049 &     9.0 &     1.2 &  -0.2794 &   0.1309 &   6.3 &   4.6 &   7.3 \\
  3.630386 &   0.000060 &     6.1 &     1.0 &   0.0584 &   0.1599 &   4.4 &   3.8 &   5.5 \\
  3.665700 &   0.000048 &     9.8 &     1.2 &   0.2405 &   0.1268 &   6.8 &   4.8 &   7.7 \\
  3.837240 &   0.000055 &     7.4 &     1.1 &   0.4720 &   0.1449 &   5.3 &   4.3 &   6.3 \\
  3.882529 &   0.000047 &    10.2 &     1.3 &   0.0320 &   0.1247 &   7.1 &   5.2 &   7.9 \\
  3.887907 &   0.000059 &     6.2 &     1.0 &   0.3588 &   0.1573 &   4.7 &   4.0 &   5.5 \\
  4.218660 &   0.000057 &     6.7 &     1.0 &  -0.3167 &   0.1522 &   5.2 &   4.4 &   5.8 \\
  4.237851 &   0.000059 &     6.3 &     1.0 &  -0.4493 &   0.1573 &   4.9 &   4.2 &   5.6 \\
  4.499595 &   0.000060 &     6.3 &     1.0 &  -0.0433 &   0.1576 &   5.1 &   4.4 &   5.6 \\
  5.129404 &   0.000061 &     6.1 &     1.0 &  -0.1767 &   0.1606 &   5.4 &   4.8 &   5.4 \\
  5.711336 &   0.000036 &    21.3 &     2.0 &   0.2129 &   0.0951 &  17.8 &  16.7 &  13.1 \\
  5.715609 &   0.000055 &     7.4 &     1.1 &   0.0737 &   0.1450 &   6.5 &   6.2 &   6.2 \\
  5.955360 &   0.000056 &     7.1 &     1.0 &  -0.4945 &   0.1482 &   6.5 &   6.2 &   6.1 \\
  5.970498 &   0.000055 &     7.3 &     1.1 &   0.0316 &   0.1458 &   6.6 &   6.4 &   6.2 \\
  6.087356 &   0.000050 &     9.0 &     1.2 &   0.0992 &   0.1314 &   8.0 &   7.9 &   7.2 \\
  6.088370 &   0.000061 &     5.9 &     1.0 &   0.4130 &   0.1627 &   5.6 &   5.4 &   5.3 \\
  6.092044 &   0.000061 &     6.0 &     1.0 &   0.4547 &   0.1625 &   5.5 &   5.4 &   5.3 \\
  7.295843 &   0.000055 &     7.3 &     1.1 &  -0.4786 &   0.1448 &   7.2 &   7.1 &   6.2 \\
 15.176058 &   0.000060 &     6.1 &     1.0 &   0.0943 &   0.1595 &   6.7 &   6.8 &   5.5 \\
 15.228742 &   0.000052 &     8.1 &     1.1 &   0.0935 &   0.1381 &   8.6 &   8.8 &   6.7 \\
 18.421901 &   0.000056 &     7.0 &     1.0 &  -0.2518 &   0.1488 &   7.4 &   7.7 &   6.0 \\
\end{longtable}

\end{appendix}

\end{document}